\documentclass[review,3p,10pt]{elsarticle}

\usepackage{amssymb}
\usepackage{amsmath}
\usepackage{graphicx}
\usepackage{epstopdf}
\usepackage{bm}
\usepackage{epsfig}
\let\vec\bm
\usepackage{color}
\usepackage{algorithm}
\usepackage{algorithmic}
\newtheorem{remark}{Remark}

\usepackage[list=true]{subcaption}
\newcommand{\ignore}[1]{}
\newcommand{\q}[1]{``#1''}
\DeclareMathOperator*{\argmin}{arg\,min}

\begin{document}

\begin{frontmatter}




\title{Efficient Propagation of Uncertainties in Manufacturing Supply Chains: Time Buckets, L-leap and Multilevel Monte Carlo}


\author{Nai-Yuan Chiang}
\ead{chiangn@utrc.utc.com}

\author{Yiqing Lin}
\ead{liny@utrc.utc.com}

\author{Quan Long\corref{mycorrespondingauthor}}
\cortext[mycorrespondingauthor]{Corresponding author. Authors are listed alphabetically.}
\ead{longq@utrc.utc.com}

\address{United Technologies Research Center\\ 411 Silver Lane, East Hartford, CT, USA}

\begin{abstract}
Uncertainty propagation of large scale discrete supply chains can be prohibitive when a large number of events occur during the simulated period and discrete event simulations (DES) are costly. We present a time bucket method to approximate and accelerate the DES of supply chains. Its stochastic version, which we call the L(logistic)-leap method, can be viewed as an extension of the leap methods, e.g., $\tau$-leap \cite{gillespie2001approximate}, $D$-leap \cite{BAYATI20095908}, developed in the chemical engineering community for the acceleration of stochastic DES of chemical reactions. The L-leap method instantaneously updates the system state vector at discrete time points and the production rates and policies of a supply chain are assumed to be stationary during each time bucket. We propose to use Multilevel Monte Carlo (MLMC) to efficiently propagate the uncertainties in a supply chain network, where the levels are naturally defined by the sizes of the time buckets of the simulations. We demonstrate the efficiency and accuracy of our methods using four numerical examples derived from a real world manufacturing material flow. In these examples, our multilevel L-leap approach can be faster than the standard Monte Carlo (MC) method by one or two orders of magnitudes without compromising the accuracy. 
\end{abstract}

%The first three examples demonstrate the performance of the time bucket method in both deterministic and stochastic settings for a simple push system, a simple pull system and a complex pull system with features like back-ordering, priority-production and transportation. The fourth example considers multilevel uncertainty propagation of the push system under parametric uncertainties. The fifth example considers multilevel uncertainty propagation of the complex pull system under both parametric and stochastic uncertainties.


\end{frontmatter}


%\section{Introduction}
%\label{}



%%  \bibliography{<your bibdatabase>}%


\section{Introduction}
Supply chains are coordinated flows of materials from the suppliers to the locations where they are supposed to be consumed. As one of the major supply chain simulation methodologies, DES concerns the modeling of a system as it evolves over time by a representation in which the state variables change instantaneously at distinct points in time \cite{law2014simulation}.  The method is commonly used to analyze complex processes that are challenging with closed-form analytical methods. DES is widely used for supply chain management analysis such as manufacturing process and logistics planning \cite{JAHANGIRIAN20101, lu2012modeling, ziarnetzky2014simulation}. Simulations enable the design of the supply chain, and the evaluation of supply chain management prior to implementation of the system to perform what-if analysis \cite{thierry2008}.

A DES model is rarely run only once. Multiple simulation runs are usually required for various purposes. As input parameters, e.g., processing time of a product, are often random variables, multiple runs with different realizations of the random input variables are required in order to obtain statistically meaningful outputs. Furthermore, if a sensitivity analysis is applied on a simulation model to select input variables that have the largest impact on response variables, another layer of multiple runs are needed to vary input parameters such as different distributions of processing times \cite{li1989marked, montevechi2007application}. Optimization is another technique that can be combined with DES to define optimal input control variables, e.g., production capacity. Each iteration of an optimization requires multiple simulation runs for a set of system parameters \cite{fu2014simulation, kapuscinski1999optimal, rosenblatt1993combined, yan2007job}.

In summary, a large amount of DES runs are often required for an analysis task. As the scale of supply chains grows large, for example, due to globalization and inter-enterprise collaboration\cite{baldwin2012global, simatupang2002collaborative}, some simulation models may take hours to complete one run. Therefore, the time to perform analysis with thousands, sometimes hundreds of thousands, of DES runs for a complex supply chain can be prohibitively long when standard MC is used. 

As an approximation of DES, the full simulated time can be divided into periods of given time buckets, $\Delta t$. Time bucket based simulation does not model the occurrence of each event, instead, it counts the number of events happening in each time bucket, at the end of which the system state is updated using the model equations. Therefore, in this approach, events can be considered to occur instantaneously at the beginning of a period \cite{thierry2008}. Note that our terminology-\q{time bucket} is consistent with part of the supply chain literature, e.g., \cite{thierry2008}, while $\Delta t$ can be equivalently denoted by \q{time interval}, \q{time leap}, etc. The size of the time bucket can be defined either as a fixed value or in a time-dependent fashion. When the size of a time bucket is small enough that each bucket has at most one event, then the model is equivalent to DES. The advantage of the time bucket method is that it is more scalable compared with DES when the size of a time bucket is relatively large. The disadvantage is that due to the aggregation of multiple events, some interactions between events are lost, thus the model is not as accurate as DES, and is less commonly used. The Tau-leap method \cite{gillespie2001approximate, cao2006efficient, moraes2014hybrid, chatterjee2005binomial} is essentially a stochastic time bucket method that has been widely used to accelerate the simulations of chemical reactions modeled by continuous time Markovian processes. Rather than simulating every discrete event, Tau-leap method simulates the stochastic change of the system states at discrete time points using constant propensity function to simulate the number of processes happening during a time bucket. Although the simulation results are biased due to the time buckets, significant acceleration can be achieved under acceptable tolerance. Recently, the D-leap method has been proposed to accelerate the simulations of delayed chemical reactions \citep{BAYATI20095908} by introducing a queue of reactions to take account of the delays. 

Our innovations are as follows. First, we extended the D-leap method to consider features of manufacturing supply chain and logistic networks in operational research. The resulting {\it{L(logistic)-leap}} method is able to consider production time, transportation time, limited capacity, pull system and priority production. Secondly, we used MLMC method based on time buckets to propagate the uncertainties in a supply chain, where most of the computational work is shifted from the expensive models, e.g., DES, to the cheap models defined by large time buckets. The proposed approach is able to match the model accuracy of DES while overcoming its scalability limitation with the help of MLMC. To the best of our knowledge, it is the first time, this type of leap method and multilevel Monte Carlo being used in the supply chain management, which opens door to more applications associated with operations research.

Section \ref{sec:lr} is a literature review of the DES and leap methods. Section \ref{sec:DES} describes the accelerated approximation of DES using the time bucket method and the detailed algorithms for the simulation of supply chain features. Section \ref{sec:DESS} introduces the L-leap method which specifically is a time bucket method for simulating logistic systems driven by stochastic processes. Section \ref{sec:UQ} presents  an MLMC method in which the samples are drawn from populations simulated using different sizes of time buckets. In Section \ref{sec:example} we show the accuracy and gain in computational speed using extensive examples. The first example concerns push system where the production does not depend on orders. The second example is a pull system with mixed orders of the spare parts and the final products, which also considers transportation delays. The third example considers the uncertainty propagation of the push system under parametric uncertainties. The fourth example considers the uncertainty propagation of the pull system under both parametric uncertainties and those driven by stochastic processes. The quantities of interests are the final delivery time of fixed amount of orders and the number of deliveries over a specified time period. We show that the error of the predictive simulations with respect to (w.r.t.) the true solution provided by DES diminishes as we decrease the size of the time bucket. We achieve a factor of several magnitudes speed up in computing the expected quantities of interest using the MLMC method based on the time buckets and L-leap, against the standard MC sampling.  

\section{Literature review}\label{sec:lr}
\subsection{Discrete event simulation in logistics and supply chains}
DES is widely used in the logistics and supply chain management as a tool to simulate the change of system states over interested time period, for example, it has been used in supply chain network structures \cite{alfieri1997object,bhaskaran1998simulation, byrne2006impact}, inventory management \cite{angulo2004supply,beamon2001performance,biswas2004object, ceroni2002workflow, dong2005performance, dong2005impacts, fleisch2005inventory, fleischmann2003integrating} and supplier selection \cite{ding2005simulation,ding2006simulation, jain2005evaluation}, etc (see \cite{tako2012application} for a detailed survey on the application of DES in the context of logistics and supply chains). In DES, the system states change instantaneously at discrete time points when relevant events take place. While the definition of events is subjected to the goal of the modeling, systematic approaches can be followed to design such a simulation \cite{law1991simulation}. The dominant type of DES is next-event time-advance where the time clock always leaps to the most imminent time among the times of occurrence of future events in an event list. The simulation complexity of a DES is therefore proportional to the number of events in the real system during a simulated period of time. Distributed computation can be used to accelerate a DES. Specifically, implementation of numerical operators, such as the random number generator and the manipulation of event list, can be parallelized. A network can be decomposed into several sub-networks whose simulations can be parallelized. Many articles have been devoted to these topics, detailed surveys can be found in \cite{fujimoto1990parallel,mustafee2014review,fujimoto2017parallel,taylor2019distributed}. 

\subsection{Time bucket method}
In time bucket method, the system clock leaps at  fixed time bucket and the system state only changes instantaneously at the end of each time bucket considering all the events occurring during the corresponding time bucket. Time bucket method can be viewed as a special case of next-event time-advance DES \cite{law2014simulation, thierry2008}. However, the procedure and analysis of time bucket method have been rarely elaborated in the literature of operational research. 
\subsection{$\tau$-leap method for the approximation of DES in chemical and biochemical systems}
$\tau$-leap method \cite{gillespie2001approximate} is a widely used time-bucket method in the simulation of discrete chemical reactions. Rather than advancing the system clock to the next time instance when a reaction process takes place (Gillespie algorithm \cite{gillespie1976general}), $\tau$-leap predict the number of reactions in a time interval using a random variable. 
\begin{align}
\Delta c_p(t+\tau) =  Poi(\tau \times r(t))\,, \label{eq:tauleap}
\end{align}
where $\Delta c_p(t+\tau)$ represents the total number of process-$p$ happening during $[t, t+\tau)$, $Poi(\tau\times r(t))$ is a Poisson random variable with parameter $\tau\times r(t)$, $r(t)$ is the rate function evaluated at time $t$. Based on the number of happened processes, we can update the system states, e.g., the number of products. Note that if $r(t)$ changes during the time period $\tau$, the method introduces time discretization error. However,  the total simulation complexity is proportional to the number of time intervals and it could be much faster than simulating every event for given numerical tolerance. Efforts have been made to enhance the efficiency and accuracy of the original version of $\tau$-leap, for instance, efficient time interval selection \cite{cao2006efficient}, postleap checking \cite{anderson2008incorporating} and hybrid method \cite{moraes2014hybrid}. In $\tau$-leap method, the reaction products are generated instantaneously without delay after molecules collide. Its extension to delayed chemical reactions leads to $D$-leap method \cite{BAYATI20095908}.

\subsection{$D$-leap method for simulation of delayed chemical and biochemical systems}
$D$-leap \cite{BAYATI20095908} is an extension of $\tau$-leap in that it considers delayed chemical reactions. It counts the number of reactions happening during a time interval using \eqref{eq:tauleap} and the reactants are instantaneously consumed, hence the system state is updated by 
\begin{align}
x_i(t+\tau) = x_i(t) - \sum_p k_{pi}\Delta c_p(t+\tau) \quad for \quad i=1,...,n_s\,, \label{eq:d-consumption}
\end{align}
where $x_i$ is the $i^{th}$ system state, $n_s$ is the number of system states, $k_{pi}$ is the consumption of $x_i$ by a single event of reaction $p$. The earliest production time is time $t$ plus the given minimum delay of reaction $p$, while the latest finishing time of $\Delta c_p$ units of reaction $p$ is $t+\tau$ plus the given maximum delay of reaction $p$. During any time interval which overlaps with the span between the earliest production time and the latest finishing time, the possible accomplished reaction $p$, which is a fraction of $\Delta c_p(t+\tau)$, is defined by a binomial distributed random variable. Consequently, the system state is updated in a similar fashion as \eqref{eq:d-consumption}. Nevertheless, the production leads to a positive change of the number of products. This constitutes the base for our $Logistic$-leap method in logistic and supply chain context where lead time of a process is usually non-negligible. 

\subsection{Monte Carlo in supply chain management}
Monte Carlo method is widely used to propagate uncertainties of random inputs to a typical quantity of interest in a supply chain \cite{schmitt2009quantifying, deleris2005risk,wong2008supply, kim2011optimal, jung2004simulation}.  Many variance reduction techniques \cite{kleijnen2013variance, law1991simulation}, e.g., antithetic variate, control variate, have been applied together with DES to increase the statistical efficiency of the uncertainty propagation. MLMC emerged recently as a powerful sampling method to accelerate the computation of an expectation via drawing samples from a hierarchy of models \citep{giles2008,giles2015}, while control variate can be viewed as the simplest form of MLMC consists two levels \citep{giles2015}. In \citep{anderson2012,moraes2016}, multilevel Monte Carlo and $\tau$-leap are applied to the stochastic simulation of chemical reactions to achieve better scalability.

\section{Time bucket approximation of DES for supply chains}\label{sec:DES}

Supply chains transport materials from the suppliers to the places where they are consumed. The raw materials usually get consumed and transformed into some intermediate products. We define set $\mathbb{P}$ of all the parts, set $\mathbb{S}\subset \mathbb{P}$  for all the supplies of raw materials and set $\mathbb{E}\subset \mathbb{P}$  for all the final products. E.g., in the supply chain of the first numerical example (see Figure \ref{fig:supplychain1}), we have eight parts among which three are raw materials, one is the final product. Hence, $\mathbb{P}=\{{\cal{P}}_1, {\cal{P}}_2, {\cal{P}}_3, {\cal{P}}_4, {\cal{P}}_5, {\cal{P}}_6, {\cal{P}}_7, {\cal{P}}_8\}$, $\mathbb{S}=\{{\cal{P}}_1, {\cal{P}}_2, {\cal{P}}_3\}$, $\mathbb{E}=\{{\cal{P}}_8\}$. The actual supply chain can be modeled as discrete mass flows with limited capacities, i.e., the production rate of each process is bounded from above. Specifically, a supply chain can be defined by a set of $n$ processes, each of which can be described as follows 
\begin{align}
 \{\alpha_{ij}\hat{p}_{ij}| {j=1:\hat{n}_i} \}\rightarrow \{ \beta_{ik} \tilde{p}_{ik}| k=1:\tilde{n}_i\} \quad i=1,...,n \,,
\end{align}
where for each process $i$, $\hat{n}_i$ is the number of consumed parts, $\tilde{n}_i$ is the number of produced parts, $\hat{p}_{ij}$ denotes the $j^{th}$ consumed part and $\tilde{p}_{ik}$ denotes the $k^{th}$ produced part. We use $\alpha_{ij}$ and $\beta_{ik}$ as the integer weights corresponding to parts $\hat{p}_{ij}$ and $\tilde{p}_{ik}$, respectively. That is, if process $i$ happens once, it consumes $\alpha_{ij}$ units of part $\hat{p}_{ij}$ and will produce  $\beta_{ik}$ units of part $\tilde{p}_{ik}$. Note that the symbols $\hat{p}_{ij}$ and $\tilde{p}_{ik}$ are ``local'' w.r.t. process $i$. A part may have different local symbol in different process. E.g.,${\cal{P}}_4$ is locally $\tilde{p}_{11}$ in process one and on the other hand, it is $\hat{p}_{32}$ in process three of the first example. By definition, $\mathbb{P}$ contains all the parts in the system, hence we have   $\mathbb{P} = \{ \hat{p}_{ij}\} \cup \{\tilde{p}_{ik}\}$, for all $i$, $j$ and $k$. $\mathbb{S} = \{ \hat{p}_{ij}\} \setminus \{\tilde{p}_{ik}\}$ and $\mathbb{E} = \{ \tilde{p}_{ij}\} \setminus \{\hat{p}_{ik}\}$ for all $i$, $j$ and $k$. We denote $\vec x \in \mathbb{R}^{|\mathbb{P}|}$ as the state vector recording the number of parts, where $|\cdot|$ denotes the set cardinal. Note that the mapping $\{ x\} \rightarrow \mathbb{P}$ is bijective, where $\{x\}$ is the set of the components of $\vec x$. Based on the definitions of $\hat{p}_{ij}$ and ${\tilde{p}_{ik}}$, we have $x_{\hat{p}_{ij}}$ as the number of the $j^{th}$ part consumed in the $i^{th}$ process, and, similarly, $x_{\tilde{p}_{ik}}$ is the number of the $k^{th}$ part produced  in the $i^{th}$ process. For clarity, in the following texts we use $\hat{x}_{ij}$ and $\tilde{x}_{ik}$ to denote these quantities. At time $t$, the process occurs at a rate $\lambda_i(t)$ which is given by  
\begin{align}\label{eq:det_lambda}
\lambda_i(t) = 
\left\{ 
\begin{array}{ll}
\displaystyle \lambda_i^{max} & \displaystyle if \quad \min_j\{\hat{x}_{ij}(t)-\alpha_{ij}\lambda_i^{max}\Delta t\} \geq 0\\
\quad&\quad \\
\displaystyle \min_j \left\{ \frac{\lfloor\frac{\hat{x}_{ij}(t)}{\alpha_{ij}}\rfloor}{\Delta t} \right\} & otherwise
\end{array},
\right.
\end{align}
where $\lambda_i^{max}$ is the maximum production rate (capacity) associated with the $i^{th}$ process, $\Delta t$ is the size of the time bucket, $\lfloor{x}\rfloor = \max\{m \in \mathbb{Z} | m \leq x\}$ is the floor function, which rounds down $x$ to the nearest integer. The first equation in \eqref{eq:det_lambda} shows that the process can achieve its maximum rate if all its materials have enough inventory in this time bucket, otherwise, the rate $\lambda_i$ is reduced to the value which prevents negative values of the consumed materials during this time bucket.  Equation \eqref{eq:det_lambda} denotes a deterministic production rate, while other alternatives are possible. For example, the consumption rate $\lambda_i$ in \eqref{eq:det_lambda} can be modeled by incorporating the expected arrivals of the consumed parts \cite{d2010modeling}, i.e., when one part, e.g., $\hat{p}_{ij}$, is out of stock, its availability in the next time bucket may be estimated by checking the scheduled productions in the preceding processes over this time bucket. If the number of scheduled productions plus the current inventory is larger than $\lambda_i^{max}\Delta t$, the maximum capacity, $\lambda_i^{max}$, can still be achieved. Otherwise, the consumption rate $\lambda_i$ can be adjusted to match the summation of the expected arrival of $\hat{p}_{ij}$ and its current inventory. However, we use \eqref{eq:det_lambda} in our approach since it is more likely preventing the negative inventory value of $\hat{p}_{ij}$. 

It is also worth mentioning that if a single part can be consumed by multiple processes, we need to define a distribution policy among the processes. In this case, one way to modify equation \eqref{eq:det_lambda} is as follows
\begin{align}\label{eq:det_lambda_alt}
\lambda_i(t) = 
\left\{ 
\begin{array}{ll}
\displaystyle \lambda_i^{max} & \displaystyle if \quad \min_j\{\hat{x}_{ij}(t)- \sum_{\{i'~|~\exists \hat{p}_{i'j'} = \hat{p}_{ij} \}} \alpha_{i'j'}\lambda_{i'}^{max}\Delta t\} \geq 0\\
\quad&\quad \\
\displaystyle \min_j \left\{ \displaystyle\frac{\left\lfloor\displaystyle\frac{\hat{x}_{ij}(t)}{  |\{i'~|~\exists \hat{p}_{i'j'} = \hat{p}_{ij} \}| \cdot \alpha_{ij} }\right\rfloor}{\Delta t} \right\} & otherwise
\end{array}
\right.\,,
\end{align}
which assumes that part $\hat{p}_{ij}$ is evenly consumed by all the processes requiring it. 
%Since the way to compute $\lambda_i$ is problem-specific in supply chain operations, in our implementation, we rank the priority of all the processes, and then let the simulation adjust $\lambda_i$ by \eqref{eq:det_lambda} for each process according to its priority. More details about the simulation algorithm are presented in the next section.

The time bucket simulation of a supply chain process can be split into two major phases:  1) material consumption: each process consumes the necessary parts instantaneously according to its production rate - $\lambda_i(t)$. 2) delayed production:  due to the required processing time (lead time) in each process, we consider all the productions require delays after materials have been instantaneously consumed. Note that our consumption-delayed-production framework follows the modeling procedures of the D-leap method for delayed chemical reaction network simulation in \cite{BAYATI20095908}. Importantly, in the context of logistics, we enriched the D-leap method with several salient features of supply chains: transportation, order-driven production (pull system), and priority production. We describe in details the time-bucket simulation of consumption-production in Sections \ref{sec:consumption} and \ref{sec:production}.

\subsection{Consumption}\label{sec:consumption}
The consumption of parts happening in each time bucket is instantaneously taken into account at the beginning of every time bucket.  In each time bucket $\Delta t$, the total number of triggered processes $i$ reads
\begin{align}\label{eq:Delta_C}
\Delta C_i(t) = \lambda_i(t)\Delta t\,.  
\end{align}

The state vector is then updated by the following equation
\begin{align}\label{eq:updateX_C}
\hat{x}_{ij}(t) = \hat{x}_{ij}(t-\Delta t) - \alpha_{ij}\Delta  C_i(t), \quad j=1,\cdots,\hat{n}_i\,. 
\end{align}

For the sake of conciseness, we omit variable $t$ and use $\Delta C_i$ instead of $\Delta C_i(t)$ in the remainder of this paper. 

At each time point, we check if the executions of the $\Delta C_i$ processes should be completed or not, and estimate the quantity of completions. In the implementation, a queue structure is created to store the necessary information, i.e., the index of the delayed process-$d_{n_{q}}$, where $n_q = 1,...,N_q$, $N_q$ is the number of process batches in the queue, the number of the delayed processes-$Q_{n_{q}}^{delay}$, the earliest time of the production being completed-$t_{n_{q}}^{s}$, the time span between the earliest and the latest times of the production being completed-$t_{n_{q}}^{span}$. 

%For the $n_{q}$-$th$ batch of delayed productions, we denote the index of the corresponding process $i$ as $d_{n_{q}}$; the earliest time that the delayed process can be finished- $t^s_{n_{q}}$; the number of delayed production $Q_{n_{q}}^{delay}$, which equals $\Delta C_i$; the time span $t_{n_{q}}^{span}$  between the first possible production time and the last possible production.

\begin{figure}[pt]
\centering
\includegraphics[scale=0.45]{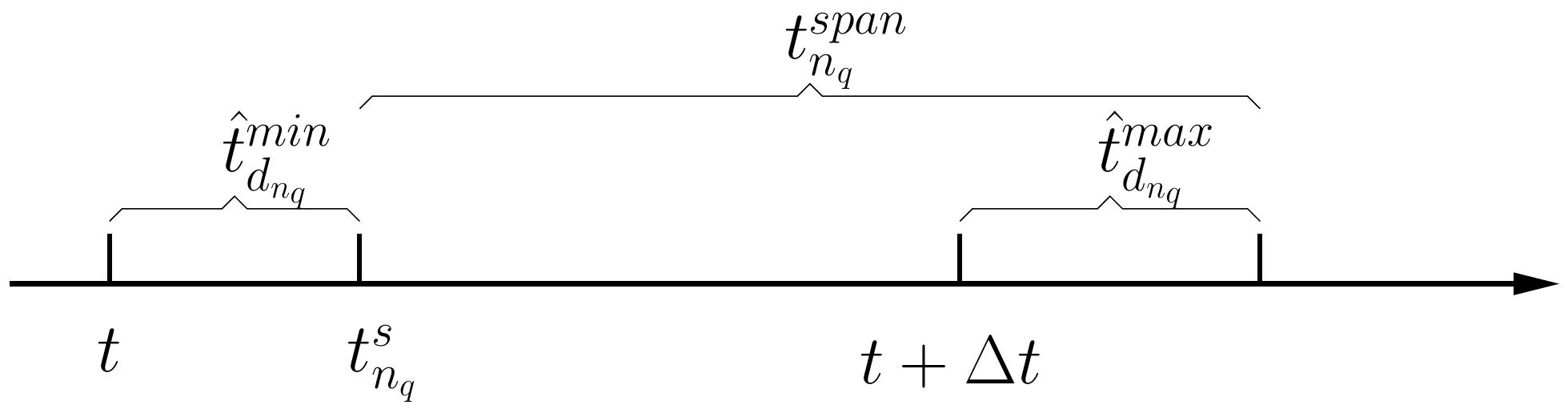}
\caption{Timeline of the processes started during $\Delta t$.}\label{fig:spanT}
\end{figure}

The earliest production time and the total production period of the $\Delta C_i$ processes can be computed as follows
\begin{align}
t_{n_{q}}^{s} = & \quad  t+ \hat{t}_{d_{n_{q}}}^{min}\,, \label{eq:compute_Ts}\\
t_{n_{q}}^{span} = & \quad t+\Delta t +  \hat{t}^{max}_{d_{n_{q}}}- t^s_{n_q} = \Delta t +  \hat{t}^{max}_{d_{n_{q}}} -   \hat{t}^{min}_{d_{n_{q}}}\,, \label{eq:compute_Tspan}
\end{align}
where $ \hat{t}^{min}_i$ and $\hat{t}^{max}_i$ are the minimum and maximum lead times for each process $i$ correspondingly. The definitions are schematically shown in Figure \ref{fig:spanT}.

We present the simulation algorithm of consumption for process $i$ in Algorithm \ref{alg:con}, which is a deterministic version of the consumption algorithm in \citep{BAYATI20095908}.
%\commentC{From line 2 to line 3, Algorithm \ref{alg:con} firstly computes the total number of consumption $\Delta C_i$ in line 2 and then updates the state vector in line 3. Lines 4-7 create intermediate variables stored in the queue structure.}

\begin{algorithm}[pht]
\caption{Consumption}\label{alg:con}
\begin{algorithmic} [1]
\STATE Input parameters: $t, \Delta t, \vec{x}, \hat{n}_i,  N_q, \lambda_i^{max},\{\alpha_{ij}\}$, $\hat{t}^{min}_i$, $\hat{t}^{max}_i$ 
%\vspace{10pt} \\
%\STATE $\Delta C_i \leftarrow 0$
\STATE    compute the total number of consumption $\Delta C_i$ using \eqref{eq:Delta_C} 
\STATE    update state vector $\{\hat{x}_{ij}\}$ using \eqref{eq:updateX_C}
%\vspace{10pt} \\
\STATE   increase the queue length by one: $ N_q \leftarrow N_q+1 $
\STATE   record the current process index in the queue structure: $d_{N_q} \leftarrow i$
\STATE   record the current consumption in the queue structure: $Q^{delay}_{N_q} \leftarrow \Delta C_i$
\STATE   compute the earliest production time $t^{s}_{N_q}$  and the production period $t^{span}_{N_q}$ using \eqref{eq:compute_Ts} and \eqref{eq:compute_Tspan}, respectively
\end{algorithmic}
\end{algorithm}

\subsection{Delayed production}\label{sec:production}
Productions are expected as long as $N_q\geq 1$. The simulation algorithm should check if there is any scheduled production due to occur in the current time bucket, i.e., all the $n_q\in\{1,\dots,N_q\}$ which satisfy $t \leq t_{n_q}^s < t+ \Delta t$. Assuming that the productions are uniformly distributed over time, the number of completed productions are proportional to the time fraction $t+\Delta t - t_{n_q}^s$ w.r.t. the total span $t^{span}_{n_q}$. Consequently, we update the associated components of the state vector-$\{\tilde{x}_{ik}\}$, $Q^{delay}_{n_q}$, $t^{s}_{n_q}$ and $t^{span}_{n_q}$ respectively.  
The details of the computations related to delayed production are summarized in Algorithm \ref{alg:prod}, which is a deterministic version of the production algorithm in \citep{BAYATI20095908}.

\begin{algorithm}[pht]
\caption{Production}\label{alg:prod}
\begin{algorithmic} [1]
\STATE Input parameters: $t$, $\Delta t$, $N_q$, $ \{\tilde{n}_i\}$,$\vec x$, $\{\beta_{ik}\}$, $\{ Q^{delay}_{n_q}\}$, $\{t^s_{n_q}\}$, $\{ t^{span}_{n_q} \}$, $\{d_{n_q}\}$ 
%\vspace{10pt} \\
%\STATE \% \% start the loop for all the entries in the queue structure \\
\FOR{$n_q \in \{1,\dots,N_q\}$ }
		\STATE get the process index from queue structure: $i \leftarrow d_{n_q}$
		%\STATE \% \% check if there is any production happend in the current time period.\\
		\IF{ $t^{span}_{n_q} > 0$ $\bf{AND}$ $t^s_{n_q} < t+\Delta t$ }
			\STATE  compute the productions happened in the current time bucket: $\Delta P_{i}  \leftarrow  Q^{delay}_{n_q} \displaystyle \min(\frac{t+\Delta t- t^s_{n_q}}{t^{span}_{n_q}},1)$  
			\STATE  update state vector: $\tilde{x}_{ik}(t) \leftarrow \tilde{x}_{ik}(t-\Delta t) + \beta_{ik}\Delta  P_i, \quad k=1:\tilde{n}_i.$
			\STATE  update queue structure: $Q^{delay}_{n_q} \leftarrow Q^{delay}_{n_q} - \Delta P_i$
			\STATE  update queue structure: $t^{s}_{n_q} \leftarrow t+\Delta t$
			\STATE  update queue structure: $t^{span}_{n_q} \leftarrow \max(0, t^{span}_{n_q}-(t+\Delta t-t^s_{n_q}))$ 
\ENDIF
\ENDFOR
\end{algorithmic}
\end{algorithm}

\subsection{Push system}
A supply chain push system, e.g., Material Requirement Planning \cite{mrp}, controls the production flow moving from the supply end to the final retailer end, with the purpose of firstly fulfilling  the raw materials in the supply end, and then starting the procedure of production according to its prediction of demands. 

Incorporating Algorithm \ref{alg:con} and  Algorithm \ref{alg:prod}, we present Algorithm \ref{alg:PushSysDet} which simulates a push system of supply chain. Note that we may need to adjust the length of the last time bucket to ensure the simulation stops at $t=T$ (lines 11-13 of Algorithm \ref{alg:PushSysDet}), where $T$ is the end time of the simulation.

\begin{algorithm}[th]
\caption{Push System of Supply Chain}\label{alg:PushSysDet}
\begin{algorithmic}[1]
\STATE Input parameters: $T$, $\Delta t$, $n$, $\{\tilde{n}_i\}$, $\{\hat{n}_i\}$, $\{\alpha_{ij}\}$, $\{\beta_{ik}\}$, $\{\lambda^{max}_i\}$, $\{\hat{t}^{min}_i\}$, $\{\hat{t}^{max}_i\}$ and $\{ \hat{x}_{ij}(0)\}$
%\vspace{10pt} \\
\STATE Initialize the queue length and the first time $t>0$: $N_q \leftarrow 0$ , $t \leftarrow \min (\Delta t, T)$ 
%\STATE \% \% start the main loop for the simulation \\
\WHILE{ $t\leq T$}
\STATE $x(t) = x(t-\Delta t)$
%\vspace{10pt} \\
%\STATE \% \%  consumption\\
\FORALL{$i \in \{1,\dots,n\}$ }
\STATE goto Algorithm \ref{alg:con} for consumption
\ENDFOR
%\vspace{10pt} \\
%\STATE \% \% production\\
\IF{ $N_q \geq 1$ }
\STATE goto Algorithm \ref{alg:prod} for productions
\ENDIF
%\vspace{10pt} \\
%\STATE \% \% adjust the time step size for the last step \\
\IF{ $t+\Delta t > T$ }
	\STATE  $\Delta t \leftarrow T - t$
\ENDIF
\STATE $t \leftarrow t + \Delta t$
\ENDWHILE
\end{algorithmic}
\end{algorithm}

\subsubsection{Inventory management}\label{sec:im}

Inventory management is usually an important part of push system. A safety stock is a popular and easy-to-implement remedy to mitigate disruptions in supply-chain operations \cite{law2014simulation, shapiro2006modeling} which can be caused by the temporal variations of product orders and the uncertainties in the supply. One strategy we can use to update the inventory is by adding the back order quantity when the inventory is less than the safety stock as follows
\begin{align}\label{eq:safety_stock}
{x}_{p}^b(t) = 
\left\{
\begin{array}{ll}
S_{p} & if~  {x}_{p}(t) \leq {x}_{p}^s \\
0 & otherwise
\end{array}
\right.\,,
\end{align}
where $p \in \mathbb{S}$ is a raw material, $x_p^s$ is the safety stock, $S_p$ is a constant used as a safeguard for the stock of part $p$. Another possible way to place the back order can be 
\begin{align}
{x}_{p}^b(t) =
\left\{
\begin{array}{ll}
 {x}_{p}^s-{x}_{p}(t) + S_{p} & if~  {x}_{p}(t) \leq {x}_{p}^s \\
0 & otherwise
\end{array}
\right. \,, \nonumber
 \end{align}
which is more resilient towards uncertainties in the supply chain network. On the other hand, when we increase the amount of inventory, we expect increased storage costs. Finding a good balance between the safety stock ${x}_{p}^s$,  safeguard $S_{p}$, order delay $t_p^d$ and costs, remains challenging in practice. The optimal strategy for inventory management is problem specific, and an extensive literature has been devoted to this topic \cite{grossmann2005enterprise,you2008mixed,daniel2005simulation,shapiro2006modeling}. 

Let $\bar{t}_p$ denote the time when the next supply of part $p$ arrives. Given a constant $M>T$, our inventory management can be summarized as in Algorithm \ref{alg:inventory} for each time $t$ when we update the system state.

\begin{algorithm}
\caption{Inventory Management}\label{alg:inventory}
\begin{algorithmic} [1]
\STATE Input parameters: $t$, $\vec x$, $\{t^{d}_p\}$, $\{\bar{t}_p\}$, $M$, $\{x^s_p\}$, $\{S_p\}$

\FORALL{$p \in \mathbb{S}$  }
	\STATE  compute the back order quantity ${x}_{p}^b$ using \eqref{eq:safety_stock}
\IF {$t\geq \bar{t}_{p}$}
	\STATE  back order arrived. Add it into the state vector: ${x}_{p} \leftarrow {x}_{p} + {x}_{p}^b$
	\STATE reset the next arrival time: $\bar{t}_{p} \leftarrow M$
\ELSIF {$ t < \bar{t}_{p}$ $\bf{AND}$  $\bar{t}_{p} =  M$ $\bf{AND}$ $x^b_p > 0$}
	\STATE  compute the next back order arrival time: $\bar{t}_{p} \leftarrow t + t_{p}^d$
\ENDIF
\ENDFOR
\end{algorithmic}
\end{algorithm}

\subsection{Pull system}
A pull system, e.g., the Toyota Production System \cite{ohno1988toyota}, for which some other names are just-in-time production and lean manufacturing, is a different policy design of manufacturing supply chains compared with a push design in that its productions and inventories managements are driven by incoming orders. In this section, we describe the time bucket algorithms for order projection before we introduce the full time bucket algorithm of pull system. The inventory management simulation should remain the same as described in section \ref{sec:im}.
%The simulation of pull system of a supply chain is more complicated than the push system, since its operation is driven by the incoming orders. Once an order has arrived, the operators need to trace back when to start the production in order to meet the deadline. Most importantly, the given supply chain must have sufficient materails at the start date, otherwise a back order from suppliers should fulfill the gap. Before introducing the full algorithm for a pull system supply chain, we firstly present some essential sub-modules, such as the inventory management and the back track logistic.

\subsubsection {Projected order and pull system}
Once a demand order is given, a supply chain system firstly check if sufficient inventory exists to meet the demand. If there is not enough inventory to fulfill the demand, the supply chain needs to start the procedure of production in order to match the gap. Hence, we need need to perform a back track to see if the existing inventories of all the intermediate parts can satisfy their own demands.   

To guarantee that all the demands are satisfied, the projected accumulated demand ${g}_p$, which includes the number of parts that is consumed in the intermediate processes, should be calculated by the following recursive function
\begin{align}\label{eq:projected_demand}
g_p(t) = 
\left\{ 
\begin{array}{ll}
\displaystyle \quad \hat{g}_p(t) & \displaystyle if \quad p\in \mathbb{E}  \\
\quad&\quad \\
\displaystyle \displaystyle \sum_{\{(i,j)| \hat{p}_{ij}=p) \}} \alpha_{ij} \max_k \left( \lceil \frac{g_{\tilde{p}_{ik}}}{\beta_{ik}}\rceil \right) + \hat{g}_p(t)
& otherwise
\end{array}
\right.\,,
\end{align}
where $\hat{g}_p = \sum_{\tau \leq t} g_p^*(\tau)$ is  the total order of part $p \in\mathbb{P}$ accumulative in time up to $t$, where $\{\tau\}$ are discrete time points in the simulation, $g_p^*(\tau)$ is the incoming order of part $p \in\mathbb{P}$ at time $\tau$, $\lceil{x}\rceil = \min\{m \in \mathbb{Z} | m \geq x\}$ is the ceiling function, which rounds up $x$ to the nearest integer. The second expression of \eqref{eq:projected_demand} consists of the direct order of part $p$ and the demand associated with the those of its ``offspring'' parts - $\sum_{\{(i,j)| \hat{p}_{ij}=p) \}} \alpha_{ij} \max_k \left( \lceil \frac{g_{\tilde{p}_{ik}}}{\beta_{ik}}\rceil \right)$.

\begin{figure}[ht]
\centering
\includegraphics[scale=0.3]{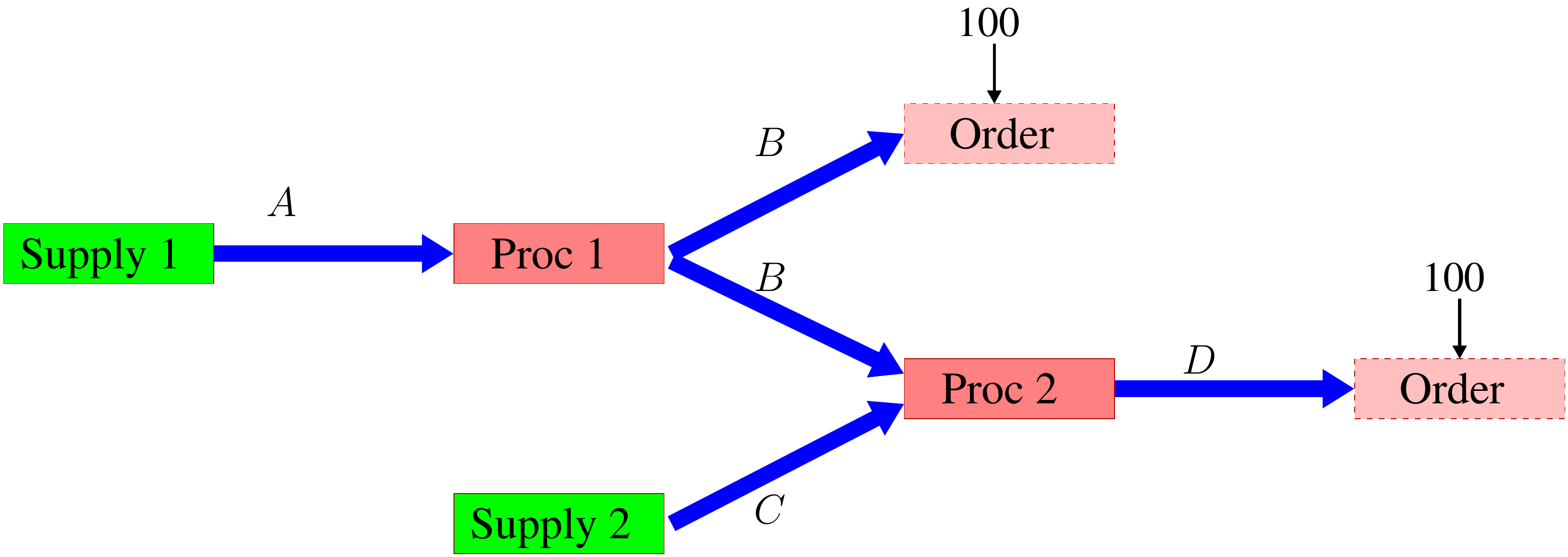}
\hspace{2cm}
\includegraphics[scale=0.3]{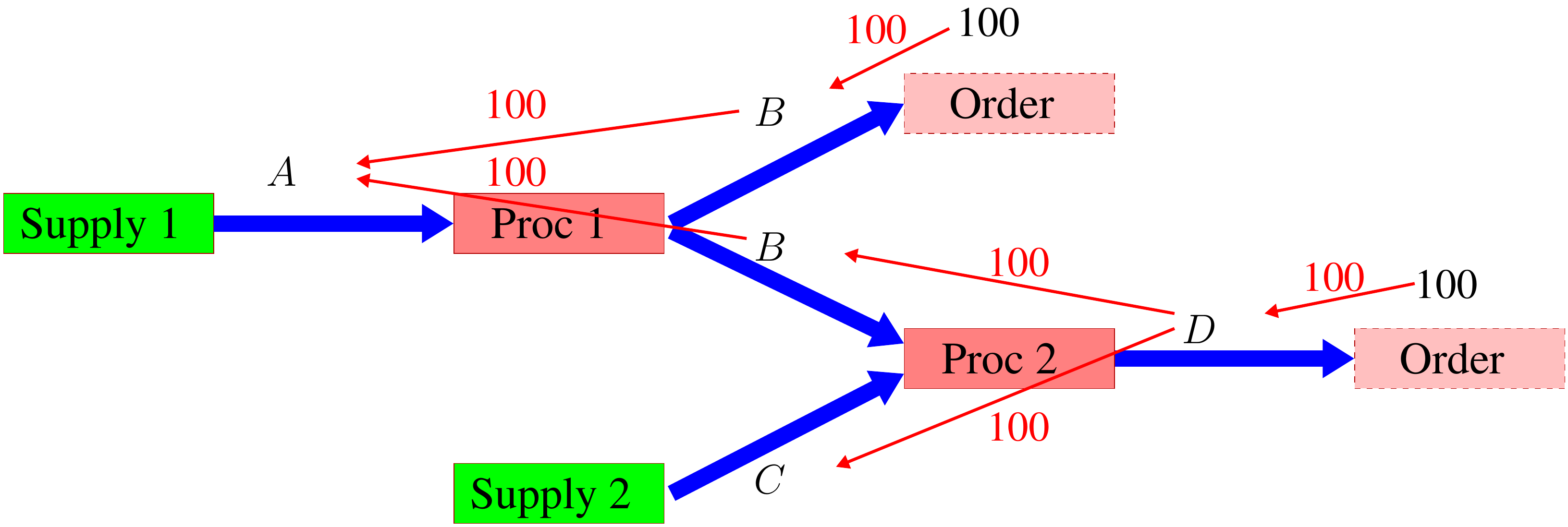}
\caption{The picture on top shows that there are two orders on part B and D, and each order requires an amount of 100 units. The bottom picture shows the projected demand of each part.}\label{fig:projectedDemand}
\end{figure}

This recursive projection can be visualized by a process starting from the final product. For example, we have a small supply chain which involves four parts as shown in Figure \ref{fig:projectedDemand}. Assuming we have some spare part orders at time $t$ on part B and D, and each order requires 100 units. By the backward recursion \eqref{eq:projected_demand}, we can obtain the projected demands for parts A, B, C and D as $200$, $200$, $100$ and $100$, respectively. 

The projected value $g_p$ indicates the necessary quantity of part $p$ that needs to be produced to satisfy the given orders. Quantity $g_p-\hat{g}_p$ then represents the least amount of part $p$ that should be consumed in the related processes. The numerical consumption may be larger than $g_p-\hat{g}_p$ for during a given $\Delta t$. In this connection, we introduce a variable $flag$ to control the consumption. The value of $flag$ is decided by comparing the accumulated consumption $c_p(t) = \sum_{\tau \leq t} \sum_{\{i | \hat{p}_{ij} =p\}}\alpha_{ij} \Delta C_i(\tau)$ with $g_p-\hat{g}_p$. On the other hand, $c_p >= g_p-\hat{g}_p$ implies that part $p$ has already been consumed sufficiently, and no more consumption should happen to it. The projected order and pull strategy is summarized in Algorithm \ref{alg:proj_pullsystem}.

%If all the necessary materials are enough in the stock, parts can be produced and this order can be satisfied eventually.  
%Otherwise, we then need to make a back order from the supplies.

\begin{algorithm}[pht]
\caption{ Projected Order and Pull Strategy}\label{alg:proj_pullsystem}
\begin{algorithmic} [1]
\STATE  Input parameters: $n$, $\{\hat{g}_p\}$, $\{c_p\}$,  $\{\alpha_{ij}\}$, $\{\beta_{ik}\}$ 
\FORALL{$i  \in \{1, \dots, n\}$ }
\STATE  $flag_i \leftarrow 0$
\ENDFOR
\FORALL{$p \in \mathbb{P}$ }
\STATE compute the projected accumulated demand $g_p$ using \eqref{eq:projected_demand}
\IF{ $c_p < g_p-\hat{g}_p$ }
	\FORALL{$ i \in \{i ~ | ~ \exists \hat{p}_{ij}=p\}$ }
		\STATE  the process that products $p$ still needs to be continued by setting $flag_i \leftarrow 1$
	\ENDFOR
\ENDIF
\ENDFOR
\end{algorithmic}
\end{algorithm}

\begin{algorithm}
\caption{Time Bucket Simulation of Supply Chain-Pull System}\label{alg:PullSysDet}
\begin{algorithmic}[1]
\STATE Input parameters: $T$, $\Delta t$, $n$, $\{\tilde{n}_i\}$, $\{\hat{n}_i\}$, $\{\alpha_{ij}\}$, $\{\beta_{ik}\}$, $\{\lambda^{max}_i\}$, $\{\hat{t}^{min}_i\}$, $\{\hat{t}^{max}_i\}$, $\{ \hat{x}_{ij}(0)\}$, $\{t^{d}_p\}$, 
$\{\bar{t}_p\}$, $M$, $\{x^s_p\}$, $\{S_p\}$, $\{\hat{g}_p\}$
\STATE Initialize the queue length $N_q \leftarrow 0$; time step size  $t \leftarrow \min (\Delta t, T)$ ; next back order arrival time $\{\bar{t}_p\} \leftarrow M$; accumulated consumption $\{c_p\} \leftarrow 0$
\WHILE{ $t \leq T$}
\STATE $x(t) = {x}(t-\Delta t)$
\STATE  goto Algorithm \ref{alg:inventory} to compute back order
\STATE  goto Algorithm \ref{alg:proj_pullsystem} to compute the projected order quantities
%\vspace{10pt} \\
\FORALL{$i \in \{1,\dots,n\}$ }
\IF{ $flag_i > 0$}
\STATE goto Algorithm \ref{alg:con}  for consumption
\FORALL{ $j \in \{1,\dots,\hat{n}_i\}$  }
\STATE  update the accumulated consumption: $ c_{\hat{p}_{ij}} \leftarrow c_{\hat{p}_{ij}} + \alpha_{ij} \Delta C_i$
\ENDFOR
\ENDIF
\ENDFOR
%\vspace{10pt} \\
\IF{ $N_q \geq 1$ }
\STATE goto Algorithm \ref{alg:prod}  for productions
\ENDIF
%\vspace{10pt} \\
\IF{ $t+\Delta t \geq T$ }
	\STATE  $\Delta t \leftarrow T - t$
\ENDIF
\STATE $t \leftarrow t + \Delta t$
\ENDWHILE
\end{algorithmic}
\end{algorithm}

\begin{remark}
The simulation using the proposed algorithms (Algorithm \ref{alg:PushSysDet} and Algorithm \ref{alg:PullSysDet}),  approaches the results from DES, when the time interval $\Delta t$ is small enough such that each individual event is resolved in the simulation. 
\end{remark}

\subsection{Hybrid system}
A hybrid system is a combination of push and pull strategies \cite{hodgson1991optimal1,hodgson1991optimal2, geraghty2004comparison, ghrayeb2009hybrid}. In a hybrid system, some of the production stages are organized by push strategies due to low level of uncertainty of the demand from their following stages, the production at the other stages, e.g., final assembly, is organized by pull strategy due to a high level of demand uncertainty. The corresponding time bucket implementation would be straightforwardly combining the push and pull strategies described in previous sections on a system level.

\section{Stochastic time bucket method: L-leap}\label{sec:DESS}

In the previous sections we presented the deterministic time bucket approximation of DES, where the number of processes happening during a fixed time interval is a deterministic value, i.e., $\Delta C_i = \lambda_i (t)\Delta t$, $i=1,\dots, n$. By introducing randomnesses into the simulation, it also allows us to have a better understanding about the potential risk in the supply-chain system. Similar to D-leap \citep{BAYATI20095908}, we treat both consumption $\Delta C_i$ and delayed production $\Delta P_i$ as random variables. Note that our framework can extend easily to the modeling of uncertainties from other sources, e.g., the demands and supplies.

We use Poisson distribution to model the number of processes happening in $\Delta t$ with parameter $\lambda_i \Delta t$ \cite{law1991simulation, BAYATI20095908}:
\begin{align}\label{eq:Delta_C_sto}
 \Delta C_i \sim Poi(\lambda_i\Delta t), \quad i = 1, \dots, n 
\end{align}

and the binomial distribution to model the number of productions \cite{law1991simulation, BAYATI20095908} in $t+\Delta t- t^s_{n_q}$ knowing the fixed number of production, $Q^{delay}_{n_q}$ during $t^{span}_{n_q}$: 
\begin{align}\label{eq:Delta_P_sto}
\Delta P_{d_{n_q}} \sim B( Q^{delay}_{n_q}, \displaystyle \min(\frac{t+\Delta t- t^s_{n_q}}{t^{span}_{n_q}},1) )\quad n_q = 1,\dots , N_q \,.
\end{align}

Note that other distributions \cite{law1991simulation} can also be possibly used to model the number of consumption and productions, which worth future research and comparison.

In addition, Algorithms \ref{alg:con} and \ref{alg:prod} can be easily extended to their stochastic version using \eqref{eq:Delta_C_sto} and \eqref{eq:Delta_P_sto}. The stochastic consumption and production can be embedded in the simulation flow of Algorithm \ref{alg:PullSysDet} which lead to a new stochastic simulation strategy. We call it the L(logistic)-leap method, where we use a constant average production rate and boolean values associated with the inventory policies and the order projections at time $t$ to predict the productions during $t$ and $t + \Delta t$. Note that compared with exact simulation of DES the approximation is used here such that we have the flexibility to accelerate the computation under prescribed numerical tolerance. Indeed, we will show that uncertainty propagation in supply chains can be dramatically accelerated without sacrificing any accuracy if we use the time bucket simulation in a coordinated way. The L-leap method we are using has a piece-wise constant rate function and its stability can be proved using the approach as described in \cite{cao2004numerical}.

%\begin{algorithm}
%\caption{L-leap Simulation of Supply Chains with Uncertainties - Pull System}\label{alg:PullSysSto}
%\begin{algorithmic}[1]
%\STATE Input parameters: $T$, $\Delta t$, $n$, $\{\tilde{n}_i\}$, $\{\hat{n}_i\}$, $\{\alpha_{ij}\}$, $\{\beta_{ik}\}$, $\{\lambda^{max}_i\}$, $\{\hat{t}^{min}_i\}$, $\{\hat{t}^{max}_i\}$, $\{ \hat{x}_{ij}(0)\}$, $\{t^{d}_p\}$, 
%$\{\bar{t}_p\}$, $M$, $\{x^s_p\}$, $\{S_p\}$, $\{\hat{g}_p\}$
%\STATE \commentC{Initialize the queue length $N_q \leftarrow 0$; time step size  $t \leftarrow \min (\Delta t, T)$ ; next back order arrival time $\{\bar{t}_p\} \leftarrow M$; accumulated consumption $\{c_p\} \leftarrow 0$}
%\WHILE{ $t \leq T$}
%\STATE $x(t) = {x}(t-\Delta t)$
%\STATE  goto Algorithm \ref{alg:inventory} \commentC{to compute back order}
%\STATE  goto Algorithm \ref{alg:proj_pullsystem} \commentC{to compute the projected order quantities}
%%\vspace{10pt} \\
%\FORALL{$i \in \{1,\dots,n\}$ }
%\IF{ $flag_i > 0$}
%\STATE goto Algorithm \ref{alg:con_sto}  \commentC{for consumptions}
%\FORALL{ $j \in \{1,\dots,\hat{n}_i\}$  }
%\STATE  \commentC{update the accumulated consumptions:} $ c_{\hat{p}_{ij}} \leftarrow c_{\hat{p}_{ij}} + \alpha_{ij} \Delta C_i$
%\ENDFOR
%\ENDIF
%\ENDFOR
%%\vspace{10pt} \\
%\IF{ $N_q \geq 1$ }
%\STATE goto Algorithm \ref{alg:prod_sto}  \commentC{for productions}
%\ENDIF
%%\vspace{10pt} \\
%\IF{ $t+\Delta t \geq T$ }
%	\STATE  $\Delta t \leftarrow T - t$
%\ENDIF
%\STATE $t \leftarrow t + \Delta t$
%\ENDWHILE
%\end{algorithmic}
%\end{algorithm}

\section{Uncertainty propagation using time bucket simulation and MLMC} \label{sec:UQ}

In this section, we describe the problem of uncertainty forward propagation, the MC discretization of an expectation, and the MLMC approach to compute the expectation. MLMC was combined with $\tau$-leap for uncertainty quantification in the context of stochastic chemical reactions in \cite{anderson2012, moraes2016}. 

Forward uncertainty propagation is concerned about the estimation of the expected value of a quantity of interest($q$), e.g., $q$ can be the delivery time of the final products. The standard MC estimator reads

\begin{align}
\bm{E}_{{\bm{\theta}}, \omega} \left[q({\bm{\theta}}, \omega)\right] = \frac{1}{N_s}\sum_{k=1}^{N_s} q({\bm{\theta}}_k, \omega_k)
+{\cal{O}_P}\left(\frac{1}{\sqrt{N_s}} \right)\,, \label{eq:exp}
\end{align}
where $\vec\theta$ is the vector of random parameters, $\omega$ is the noise which perturbs the system states dynamically, $N_s$ is the number of samples. The notation of sequence of random variables $Y_{N_s} = {\cal{O}_P}\left(d_{N_s}\right)$ indexed by $N_s$ means that for any $\epsilon > 0$, there exists a finite $K$ and a finite $N_0$, such that for any $N_s > N_0$, the probability $Pr( Y_{N_s} > K d_{N_s})$ is smaller than $\epsilon$.  Assigning a tolerance $\epsilon_s$ and a confidence level $\alpha$ on the statistical error leads to
\begin{align}
Pr\left( \left|\frac{1}{N_s}\sum_{k=1}^{N_s} q({\bm{\theta}}_k, \omega_k) - \bm{E}_{{\bm{\theta}}, \omega} \left[q({\bm{\theta}}, \omega)\right]\right| < \epsilon_s\right) 
= \alpha \label{eq:stol}
\end{align}

Considering the Central Limit Theorem (CLT), i.e., $\frac{1}{N_s}\sum_{k=1}^{N_s} q({\bm{\theta}}_k, \omega_k) - \bm{E}_{{\bm{\theta}}, \omega} \left[q({\bm{\theta}}, \omega)\right]\sim{\cal{N}}(0, \frac{V}{N_s})\quad as \quad N_s \rightarrow \infty$, we can equivalently express \eqref{eq:stol} using the distribution function of a standard normal:
\[\phi(\frac{\sqrt{N_s}\epsilon_s}{\sqrt{V}}) = \frac{1+\alpha}{2}\,.\] 
Consequently, we obtain the expected number of samples in order to control the statistical error in probability: 
\begin{align}
N_s = V \Phi^{-2}(\frac{1+\alpha}{2}) \epsilon_s^{-2}\,,
\end{align}
where $V$ is the variance of the quantity of interest, $\Phi^{-1}(\cdot)$ is the inverse distribution function of the standard normal distribution, $\epsilon_s$ is the tolerance on the absolute error committed by the MC estimator. Then, the total computational cost of a standard MC sampler is:
\begin{align}
C_{mc} = C_m V \Phi^{-2}(\frac{1+\alpha}{2}) \epsilon_s^{-2}\,, \label{eq:MC}
\end{align}
where $C_m$ is the average cost of a single DES. 
 
MLMC is optimized in the sense that the total computational cost is minimized for a given tolerance on the numerical error. In the hierarchy of models, high level models are more accurate and computationally more expensive than low level models. Provided that the expectation and variance of the difference between the approximated and the true solutions diminish at certain rates, as the level increases, we can construct an MLMC sampler, which can be several orders more efficient than the standard MC method. Note that standard MC method would put all its samples on the highest level to control the bias of the estimator. Let $q_l$ denote the corresponding level $l$ approximation of the quantity of interest $q$. Assume that the numerical discretization error is bounded uniformly in the probability space as follows
\begin{align}
\bm{E} (q - q_l) = {\cal{O}} \left( \Delta t_l^a \right)\,, \label{eq:bias1}
\end{align}
where $\Delta t_l$ is the size of the time bucket on level $l$, $a \in \mathbb{R}^+$ is the convergence rate of the numerical discretization, the notation $Y_{\Delta t} ={\cal{O}} \left( d_{\Delta t} \right)$ indexed by $\Delta t$ is the deterministic version of $Y_{\Delta t} ={\cal{O}_P} \left( d_{\Delta t} \right)$, which means that there exists a finite $K$ and a finite $\Delta t_0$, such that for any $\Delta t < \Delta t_0$, $Y_{\Delta t} \leq K d_{\Delta t}$.

The expectation in \eqref{eq:exp} can be rewritten as a telescopic sum as follows

\begin{align}
\bm{E}(q) = \sum_{l=0}^L \bm{E}(q_l -q_{l-1}) + {\cal{O}}\left(\Delta t_L^{a} \right)\,, \quad \text{with}\quad q_{-1} = 0\,. \label{eq:telesum}
\end{align}

Furthermore, we can write the first term on the right hand side (r.h.s.) of Equation \eqref{eq:telesum} as a summation of sample averages, and \eqref{eq:telesum} becomes
\begin{align}
\bm{E}(q) = \hat{q} + \sum_{l=0}^{L}{\cal{O}_P} \left( \frac{1}{\sqrt{N_l}} \right) + {\cal{O}}\left(\Delta t_L^{a} \right)\,, \quad \text{with}\quad q_{-1} = 0\,,
\end{align}
where 
\begin{align}
\hat{q} = \sum_{l=0}^L \frac{1}{N_l} \sum_{k=1}^{N_l}(q^k_l -q^k_{l-1})\,, \label{eq:mlmcestimator}
\end{align}
is the MLMC estimator of $q$, $\sum_{l=0}^{L}{\cal{O}_P} \left( \frac{1}{\sqrt{N_l}} \right)$ is the statistical error, ${\cal{O}}\left(\Delta t^a_L\right)$ is the numerical bias. A heuristic argument on the computational advantage of using this estimator is the following: the variance of $q_l - q_{l-1}$ becomes very small as $l$ increases, hence we draw few high-level samples while most of the samples are shifted to the lower levels where the computations are fast. 

Next, we optimize the computational cost of the MLMC estimator for given tolerances on the bias and the statistical error which read 
\begin{align}
\bm{E}(q - q_L) &= \epsilon_b\,, \label{eq:bias2}\\
Pr( |\hat{q} - \bm{E}(q_L) |< \epsilon_s) &= \alpha\,, \label{eq:statisticalerror}
\end{align} 
where $\epsilon_b$ is the tolerance on the bias, $\epsilon_s$ is the tolerance on the statistical error. Note that we can use CLT to convert \eqref{eq:statisticalerror} to the following variance constraint:
\begin{align}
\bm{Var}(\hat q) = \frac{\epsilon_s^2}{\Phi^{-2}(\frac{1+\alpha}{2})}\,. \label{eq:varianceControl}
\end{align}

The maximum level can be obtained from \eqref{eq:bias1} and \eqref{eq:bias2}:
\begin{align}
	L = \frac{1}{a} log_2(\epsilon_b)\,, \nonumber
\end{align}
assuming that $2^{-l} = \Delta t_l$. 

The optimal number of samples on each level can be obtained by minimizing the total cost under the constraint \eqref{eq:varianceControl} on the variance of the estimator:
\begin{align}
\{N^{opt}_l ,\, l=0,...,L\}= \argmin_{\{N_l,\, l=0,...,L\} } \left[\sum_{l=0}^L C_l N_l  + \lambda \left(\sum_{l=0}^L\frac{V_l}{N_l} -  \frac{\epsilon_s^2}{\Phi^{-2}(\frac{1+\alpha}{2})}\right)\right]\,, \nonumber
\end{align}
where $C_l$ is the average computational cost of $q_l - q_{l-1}$, $V_l$ is the variance of the random variable $q_l - q_{l-1}$, $\lambda$ is a Lagrangian multiplier (by an abuse of notation).
 
Solving the above minimization problem leads to 
\begin{align}
N^{opt}_l = \sqrt{\frac{V_l}{C_l}} \frac{\sum_{l=0}^L \sqrt{C_l V_l}}{\bar{\epsilon}^2_s}\quad \text{with} \quad
\bar{\epsilon}^2_s =  \frac{\epsilon_s^2}{\Phi^{-2}(\frac{1+\alpha}{2})}\,. \nonumber
\end{align}

Consequently, the optimal total computational cost of the multilevel estimator is 
\begin{align}
\sum_{l=0}^L C_l N^{opt}_l &= \sum_{l=0}^L C_l \sqrt{\frac{V_l}{C_l}} \frac{\sum_{l=0}^L \sqrt{C_l V_l}}{\bar{\epsilon}^2_s} \nonumber\\
&= \left(\sum_{l=0}^L \sqrt{C_l V_l} \right)^2 \bar{\epsilon}^{-2}_s\,. \nonumber
\end{align}
 
It is common that the variance $V_l$ and cost $C_l$ have the asymptotic bounds: $V_l ={\cal{O}} \left( \Delta t_l^b \right)$ and 
$C_l ={\cal{O}} \left( \Delta t_l^{-g}  \right)$, where $b$ and $g$ are the rates which describe the algebraic decrease/grow of the variances and computational costs, respectively. In the cases where $C_0 V_0 >> C_1 V_1 >>\dots >> C_L V_L$, the total cost is dominated by 
$C_0 V_0 \bar{\epsilon}^{-2}_s$. In the cases where $C_L V_L >> C_{L-1} V_{L-1} >>\dots >> C_0 V_0$, the total cost is dominated by 
$C_L V_L \bar{\epsilon}^{-2}_s=C_0V_0 \epsilon_b^{-\frac{g-b}{a}}\bar{\epsilon}^{-2}_s$. In the cases where $C_L V_L = C_{L-1} V_{L-1} =\dots = C_0V_0$, 
the total cost is $L^2 C_0V_0 \bar{\epsilon}_s^{-2} = C_0V_0 (log_2 \epsilon_b)^2\bar{\epsilon}^{-2}_s /a^2 $. Note that in the literature, it is common to impose a total tolerance $\epsilon^2$ on the mean square error of the MLMC estimator and split the error budget into two parts - $\theta \epsilon^2$ and $(1-\theta) \epsilon^2$ ($0 < \theta < 1$) on the bias and variance \citep{MG2003,giles2008,giles2015}. We give an explicit confidence level to the statistical error control in this study which is consistent to the literature such as \cite{collier2015continuation,moraes2016, BECK2018523}.

%If there exist independent estimator $Y_l$  based on $N_l$ MC samples, each with expected cost $C_l$ and variance constant $V_l$, and positive constants $a$, $b$, $g$,  
%\[|E(X_l-X)|< c_1 2^{-a l}\,,\] 
%\[E(Y_l)=E(X_0), \, l=0\,\] 
%\[E(Y_l)=E(X_l - X_{l-1}), l > 0\,\] 
%\[V_l \leq c_2 2^{-b l}\,,\] 
%\[C_l \leq c_3 2^{g l}\,.\] 
%
%
%
%We summarize the MLMC theorem \cite{giles2008, giles2015} as follows.  
%
%Let $X$ denote a random variable, and let $X_l$ denote the corresponding level $l$ numerical approximation. If there exist independent estimator $Y_l$  based on $N_l$ MC samples, each with expected cost $C_l$ and variance constant $V_l$, and positive constants $a$, $b$, $g$, $c_1$, $c_2$, $c_3$, such that $ a< 0.5 min(b, g)$ and 
%\[|E(X_l-X)|< c_1 2^{-a l}\,,\] 
%\[E(Y_l)=E(X_0), \, l=0\,\] 
%\[E(Y_l)=E(X_l - X_{l-1}), l > 0\,\] 
%\[V_l \leq c_2 2^{-b l}\,,\] 
%\[C_l \leq c_3 2^{g l}\,.\] 
%Then there exists a positive constant $c_4$ such that for any $\epsilon< e^{-1}$, there are values $L$ and $N_l$ for which the multilevel estimator 
%$Y=\sum Y_l$ has a mean square error with bound 
%\[MSE = E((Y-E(P))^2) < \epsilon^2\]
%with a computational complexity $C$ with bound 
%\[E(C_{MLMC}) = O\left( \epsilon^{-2} \right), b > g\, , \]
%\[E(C_{MLMC}) = O\left( \epsilon^{-2} (log\epsilon)^2 \right), b = g\, , \]
%\[E(C_{MLMC}) = O\left( \epsilon^{-2-\frac{\gamma -\beta}{\alpha}} \right), b < g\, . \]

\begin{remark}
In the case where
$C_L \approx C_m $, the complexity of a standard MC sampler is ${\cal{O}}\left(\bar{\epsilon}_s^{-2} \epsilon_b^{-\frac{g}{a}} \right)$, where $\bar{\epsilon}_s^{-2}$ is proportional to the number of samples of a standard MC, $\epsilon_b^{-\frac{g}{a}}$ is proportional to $C_L$ the computational cost of each sample on highest level $L$. Therefore, the computational complexity of MC \eqref{eq:MC} would always be asymptotically higher than those of the MLMC, namely ${\cal{O}}\left(\bar{\epsilon}^{-2}_s\right)$, 
${\cal{O}}\left(\epsilon_b^{-\frac{g-b}{a}}\bar{\epsilon}^{-2}_s\right)$ and ${\cal{O}}\left((log_2 \epsilon_b)^2\bar{\epsilon}^{-2}_s\right) $. 
\end{remark}

\begin{remark}
The $q^k_l$ and $q^k_{l-1}$ in \eqref{eq:mlmcestimator} should always be computed using the same realization of the random parameters as their inputs, to assure the correlation between $q_l$ and $q_{l-1}$. 
In the cases where the randomness is driven by stochastic processes, we adopt the coupling scheme proposed as the Algorithm 2 in \cite{anderson2012}. The key idea of this algorithm was to use the additivity property of Poisson processes to tightly correlate two processes on different levels. 
\end{remark}

\section{Numerical example}\label{sec:example}
We present in this section four numerical examples with increasing complexities. The first example is a manufacturing material flow simulated using deterministic and stochastic time bucket methods. The second example is a pull system considering back-ordering, priority delivery, and transportation delays simulated using time bucket methods. We carry out uncertainty propagation using MLMC in the third and fourth examples. We use MATLAB to implement the time bucket algorithm and build our code of MLMC on the original version from https://people.maths.ox.ac.uk/gilesm/mlmc/.

\subsection{Time bucket approximations of a simple push supply chain network}\label{eg:push}
\begin{figure}[pht]
\centering
\includegraphics[scale=0.3]{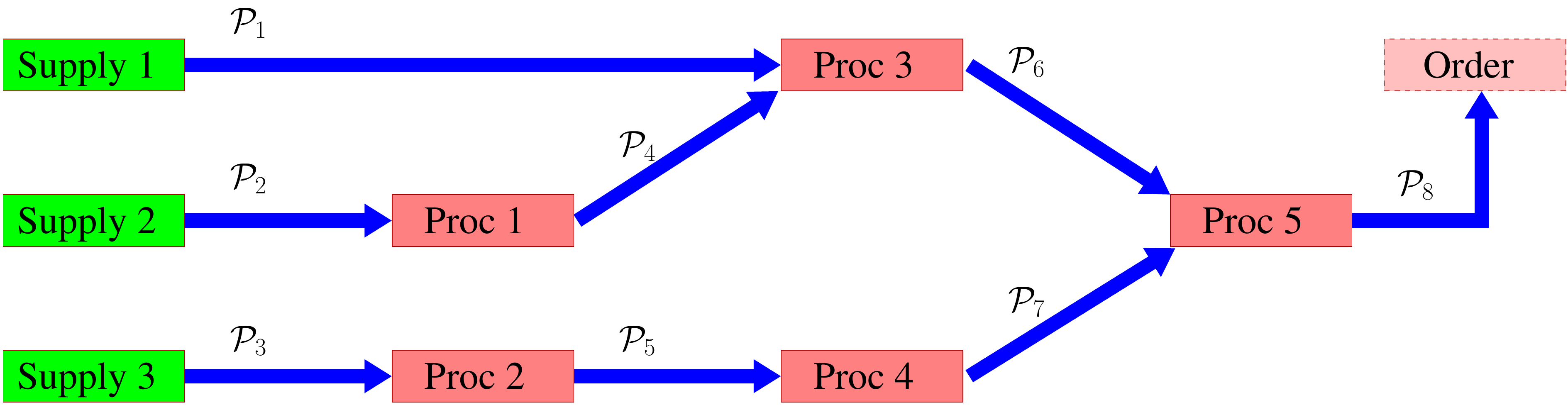}
\caption{A manufacturing system with five processes and eight parts.}\label{fig:supplychain1}
\end{figure}

We consider a supply chain system for manufacturing industry which is schematically shown in Figure \ref{fig:supplychain1}. It involves five processes and eight parts, and we show the consumption-production relationships in equations \eqref{ex1}-\eqref{ex1_end}. The parts on the left hand side of the equations are instantaneously consumed when the processes get started, while the parts on the r.h.s. are produced after certain periods of delays, characterized by the production time/lead time of each process. The production rate which describes the capacity of a process is the number of parts which get processed in a time unit, e.g., one day. 

\begin{align}
{\cal{P}}_2 = {\cal{P}}_4 \label{ex1} \\
{\cal{P}}_3 = {\cal{P}}_5\\
{\cal{P}}_1 + {\cal{P}}_4 = {\cal{P}}_6 \\
{\cal{P}}_5 = {\cal{P}}_7 \\
{\cal{P}}_6 + {\cal{P}}_7 = {\cal{P}}_8 \label{ex1_end}
\end{align}

A push system starts the procedure of production according to its prediction of demands. We assume the following initial conditions: 
$x_1(t=0)=1000$, $x_2(t=0)=500$, $x_3(t=0)=1000$, which prescribe the initial inventory levels of ${\cal{P}}_1$, ${\cal{P}}_2$ and ${\cal{P}}_3$. For all $i \notin \mathbb{S}$, i.e., the intermediate and final products, we let $x_i(t=0)=0$. Firstly we assume deterministic production rates which read $\lambda_1 = 8$, $\lambda_2=8$, $\lambda_3=4$, $\lambda_4=8$ and $\lambda_5 = 2$. We also assume the  processing time is deterministic, i.e., $\hat{t}^{max}_i = \hat{t}^{min}_i$, $i=1,\cdots,5$, and they are specifically $\hat{t}^{min}_1 = 1$, $\hat{t}^{min}_2 = 1$, $\hat{t}^{min}_3 = 10$, $\hat{t}^{min}_4 = 1$, $\hat{t}^{min}_5 = 10$.

\begin{figure}[pht]
\centering
\includegraphics[scale=0.6]{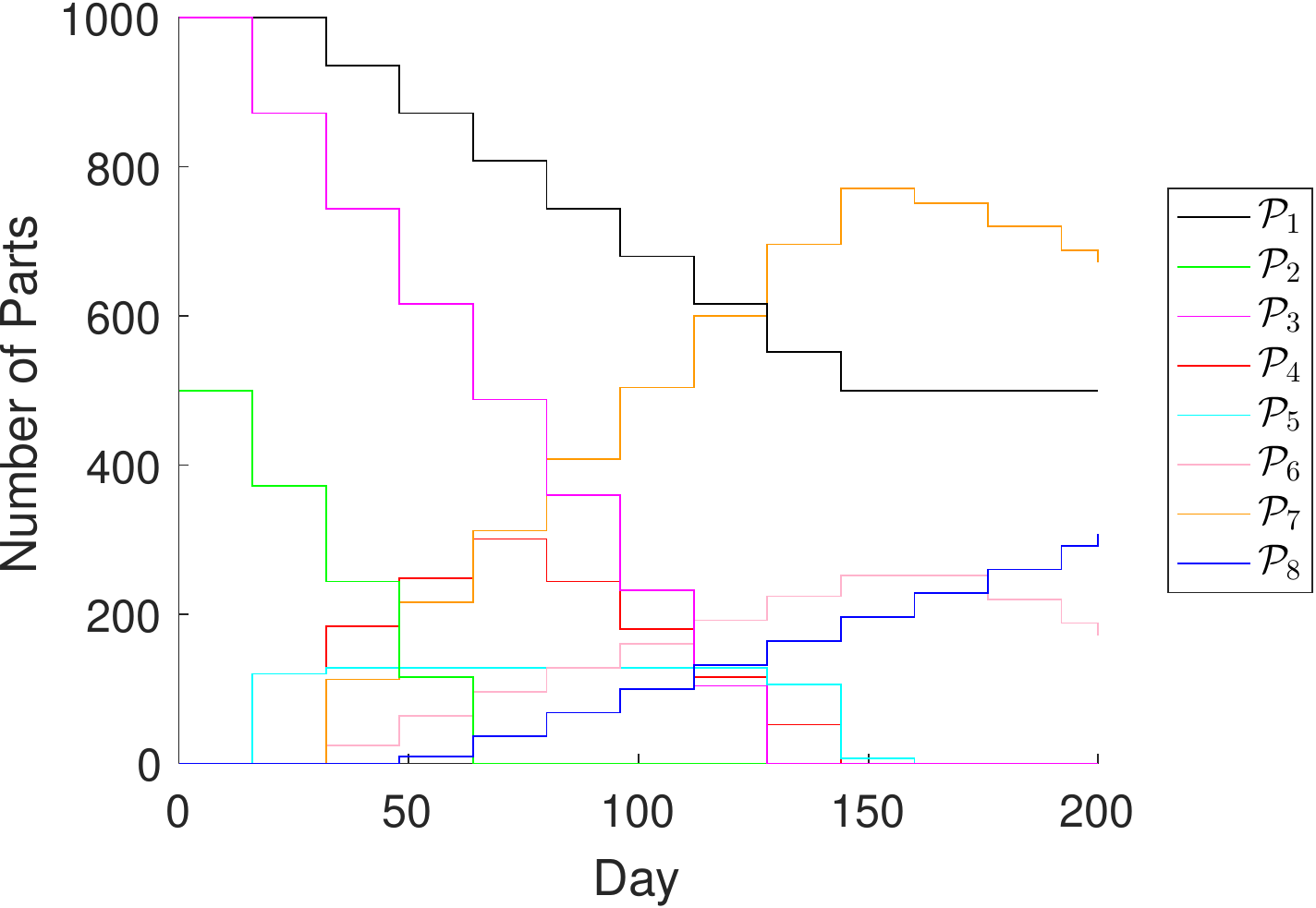}
\includegraphics[scale=0.6]{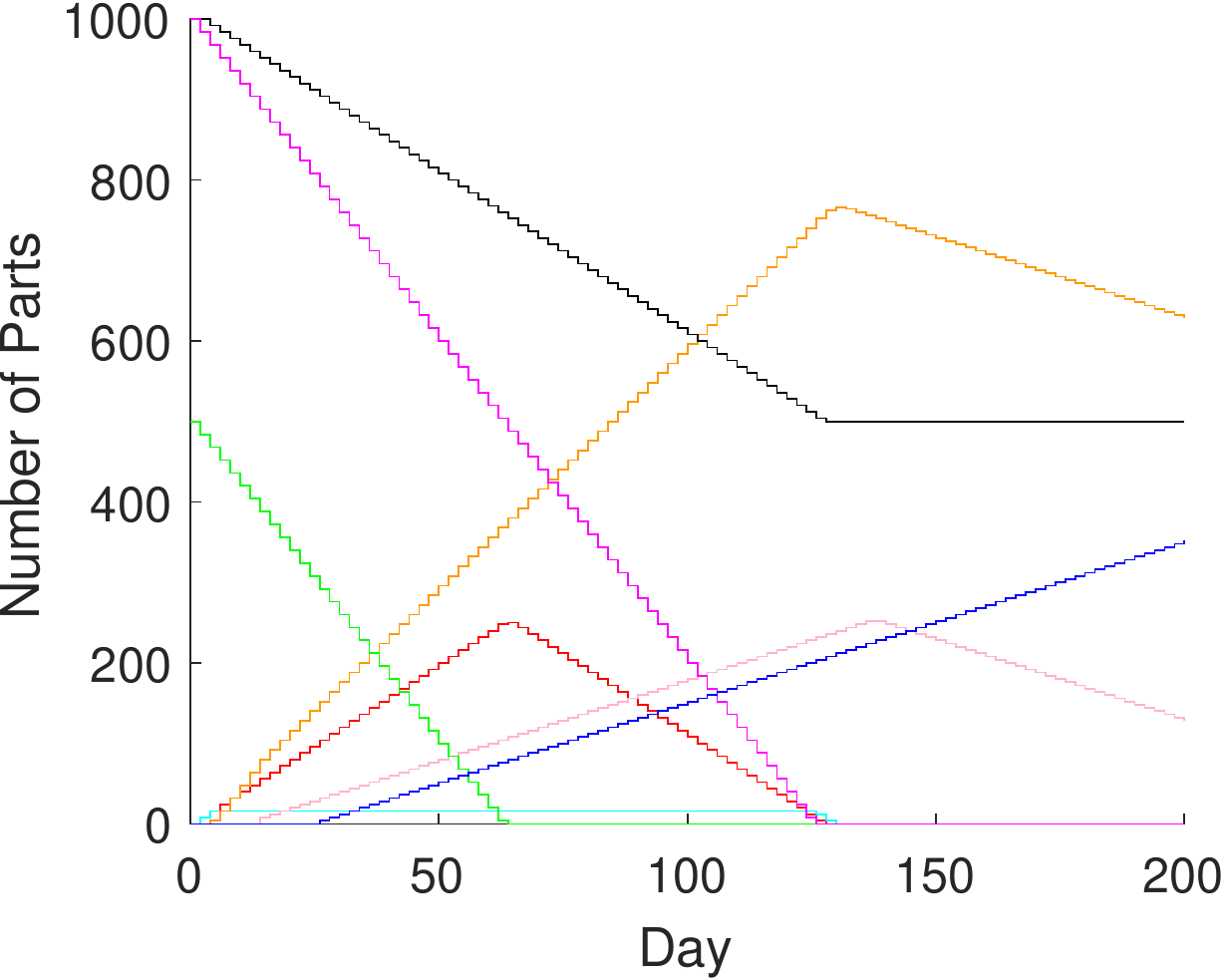}
\caption{The time history of the state vector in the push system. From left to right, figures present the case for $\Delta t = 16$ days and $\Delta t = 2$ days, respectively.}\label{fig:dt16-dt2}
\end{figure}

Figure \ref{fig:dt16-dt2} shows the time histories of the state vector, which represents the number of each part in the system at any given time, simulated under two different values of the time bucket. Note that the time bucket approximation is able to capture the main dynamical features of the system even when a coarse time bucket size, $\Delta t = 16$ days, is used. The monotonic decrease of ${\cal{P}}_1$ stops at $500$ due to the initial inventory level of ${\cal{P}}_2$. $x_8$ monotonically increases after an initial period of waiting which attributes to the production delays. The dynamics of the intermediate parts - ${\cal{P}}_4$, ${\cal{P}}_5$, ${\cal{P}}_6$ and ${\cal{P}}_7$ are majorly determined by their consumption and production rates. 

\begin{figure}[pht]
\centering
\includegraphics[scale=0.6]{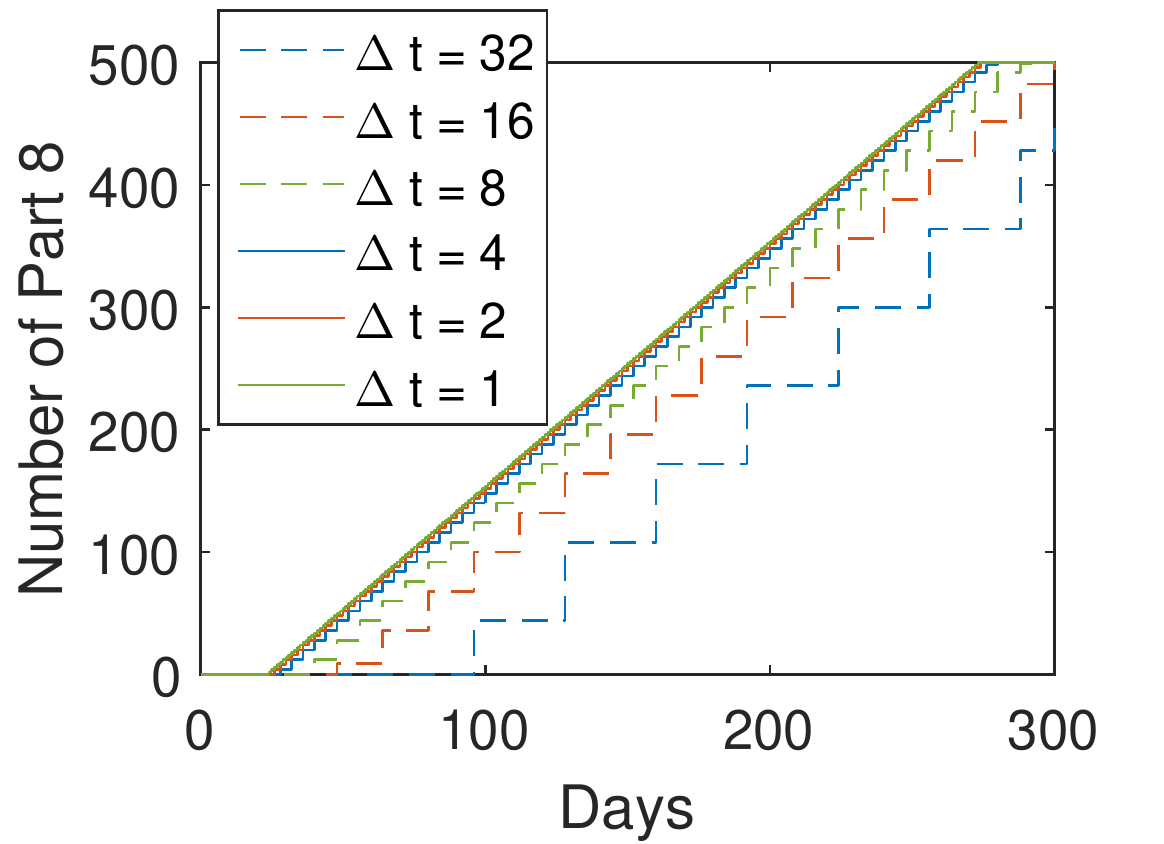}
\includegraphics[scale=0.48]{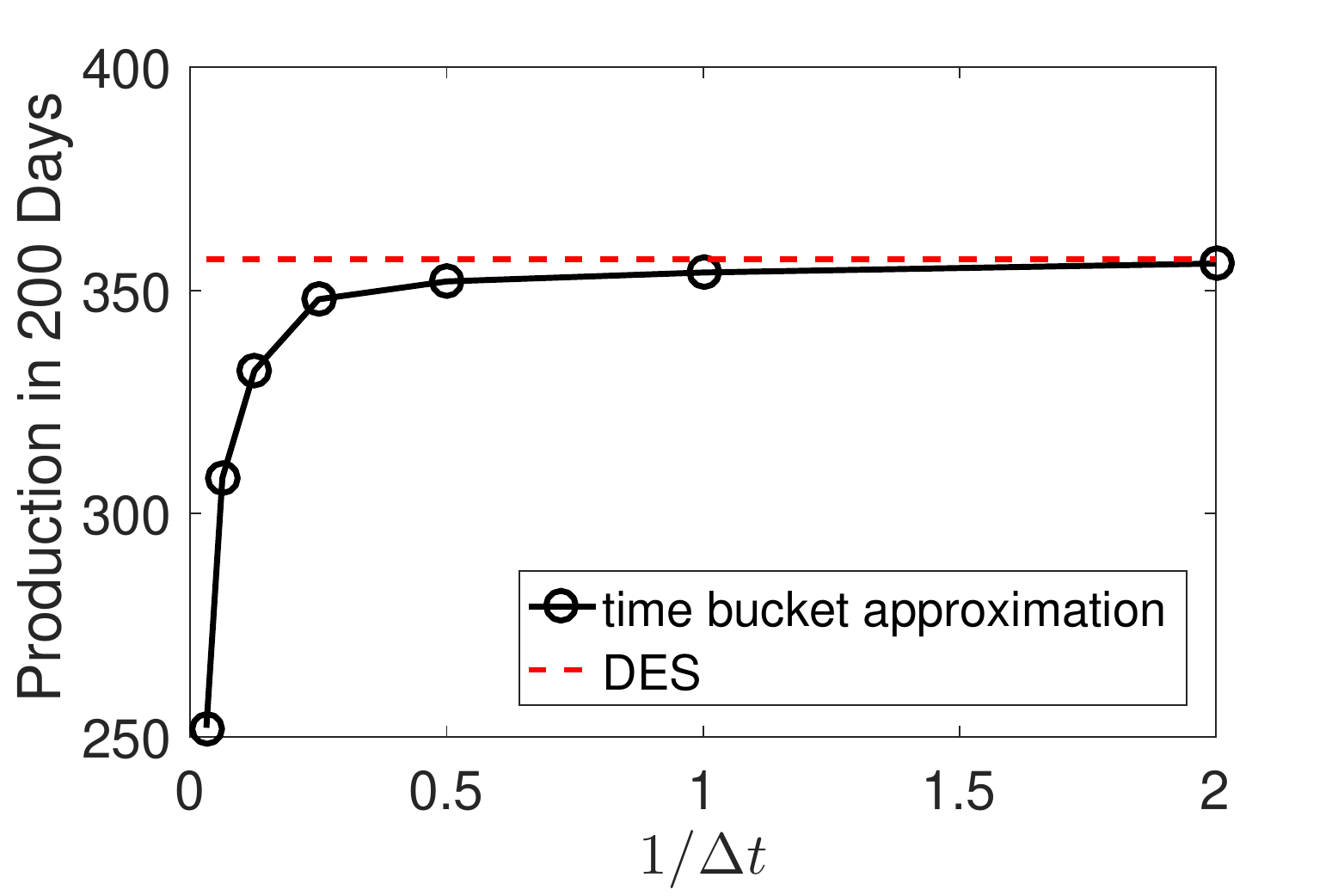}
\caption{Push system. Left figure is the simulated evolution of the number of final products (${\cal{P}}_8$); right figure is the convergence of the number of product in 200 days w.r.t. the reciprocal of the size of time bucket. The reference value is $357$.}\label{fig:convPart8}
\end{figure}

It is shown in Figure \ref{fig:convPart8} that the time bucket method converges to the ``ground truth'' computed by DES, when $\Delta t$ reduces from $32$ days to $4$ days. The error is smaller than $2\%$ when the time bucket is smaller than $4$ days. 

\begin{figure}[pht]
\centering
\includegraphics[scale=0.65]{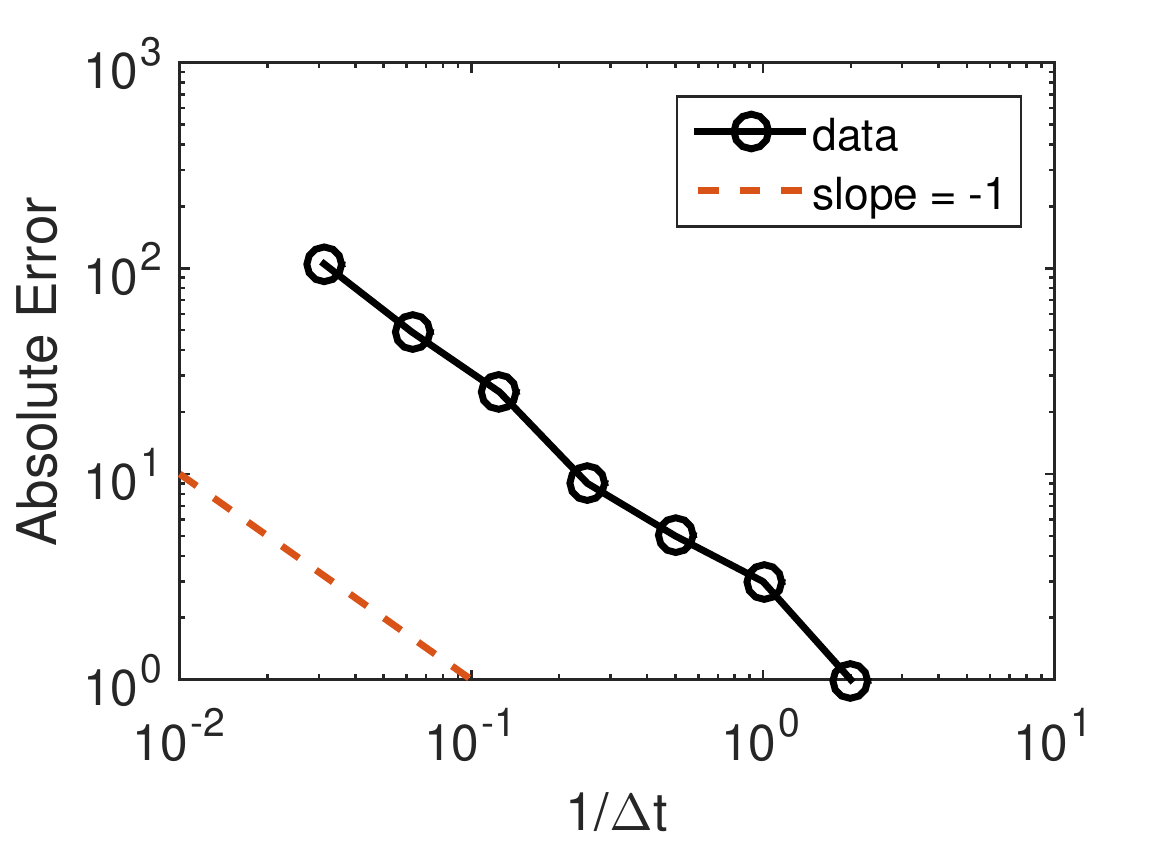}
\includegraphics[scale=0.65]{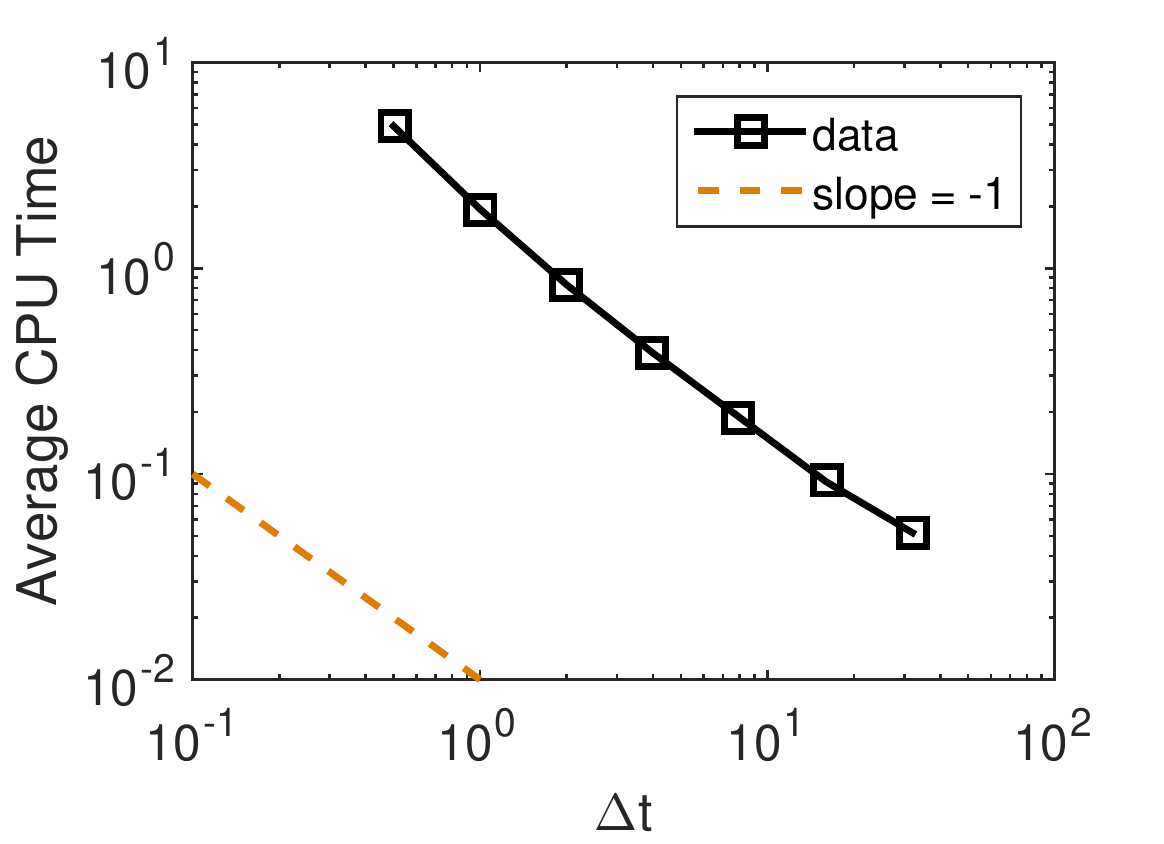}
\caption{Push system. Left figure is the absolute error of the 200 days production w.r.t. the reciprocal of the size of time bucket. Right figure is the CPU time averaged over $100$ repetitive runs of the simulation up to 200 days, w.r.t. the size of time bucket.}\label{fig:error_costPart8}
\end{figure}

The absolute error of the 200 days' production decreases linearly when time bucket size decreases as shown in the left picture of Figure \ref{fig:error_costPart8}. The CPU time of the time bucket approximation increases linearly as we increase the number of time buckets during the simulation time (The CPU time is an average value over $100$ repetitive runs). 

\begin{figure}[pht]
\centering
\includegraphics[scale=0.55]{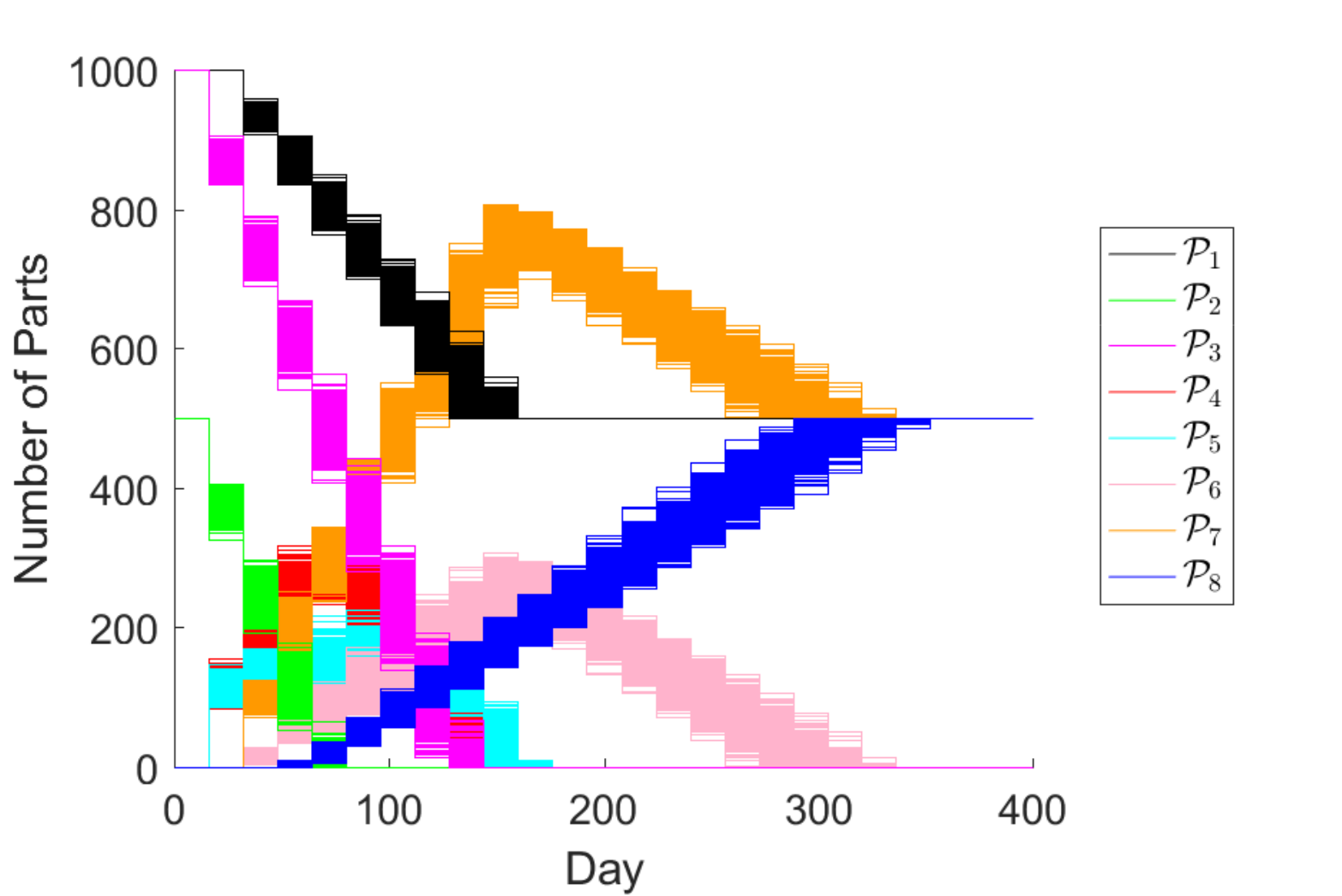}
\includegraphics[scale=0.55]{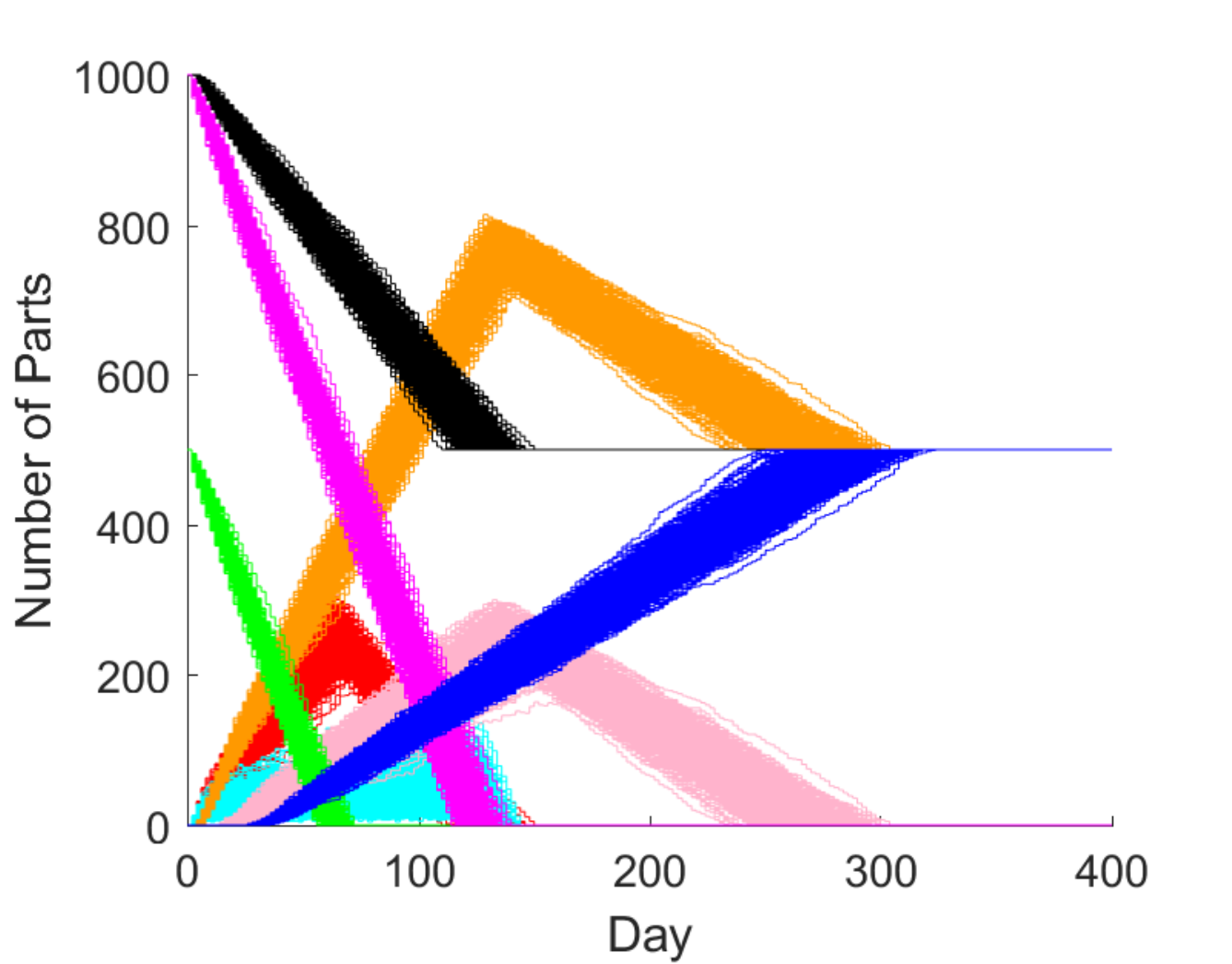}
\caption{The time history of the state vector in stochastic push system. From left to right, figures present the case for $\Delta t = 16$ days and $\Delta t = 2$ days, respectively.} \label{fig:sto-dt2-push-all}
\end{figure}

Next, we use the L-leap method to approximately simulate the stochastic system where the state vector is dynamically driven by Poisson processes. Figure \ref{fig:sto-dt2-push-all} visualizes $1000$ trajectories using identical initial data. 

In addition, we notice that the average trajectories shift from left to right when we reduce $\Delta t$, for example, the $500^{th}$ ${\cal{P}}_8$ is produced in around 300 days when $\Delta t = 16$ days, while it is produced in around 270 days with $\Delta t = 2$ days. This is due to the artificially delayed availability of its previous parts when the time bucket is coarse. 

\subsection{Time bucket approximations of a complex pull system }\label{eg:pull_t}
This example is a pull system dealing with mixed orders of spare parts and final products, and considering transportation. The system receives spare-part orders of ${\cal{P}}_4$, ${\cal{P}}_5$, ${\cal{P}}_6$ and ${\cal{P}}_7$ every $50$ days. Meanwhile, the following inventory policy is adopted to refill ${\cal{P}}_1$, ${\cal{P}}_2$ and ${\cal{P}}_3$: when the number of an inventory falls below $200$, back orders of $200$, $250$ and $300$ are placed for ${\cal{P}}_1$, ${\cal{P}}_2$ and ${\cal{P}}_3$, respectively. The delivery delays are $15$, $20$ and $30$ days for ${\cal{P}}_1$, ${\cal{P}}_2$ and ${\cal{P}}_3$, respectively. Moreover, each of them has an initial inventory of $500$. On top of the spare-part orders, we place three orders of final products on the first day, the $100^{th}$ day and the $200^{th}$ day, while each order consists $100$ final products ${\cal{P}}_8$. Upon the receipt of orders of the final products, the parts are used with priority for the production of the final products. 

\begin{figure}[h]
\centering
\includegraphics[scale=0.3]{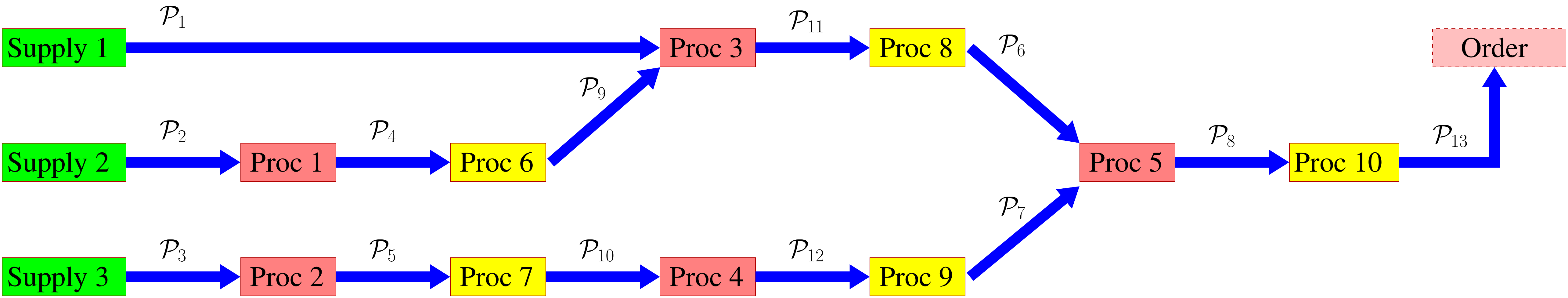}
\caption{A modified manufacturing system which includes transportation.} \label{fig:supplychain2}
\end{figure}

Additionally, transportation occurs between any two consecutive processes. The transportation of products are modeled as additional processes characterized by transportation rates and transportation delay time (similar to the production rates and the processing time of a production process). After renumerating and augmenting the original set of processes, the new system is shown in Figure \ref{fig:supplychain2}, where the first five processes are the original processes; the second set of five processes are the transportation processes. For Processes $6$-$10$, we use transportation rates $\lambda_i = 8, \forall i \in \{6,\dots,10\}$ and constant transportation delay time $\hat{t}_i^{min} = \hat{t}_i^{max}$ are  $10$, $10$, $10$, $50$ and $10$ days, respectively.

\begin{figure}[pht]
\centering
\includegraphics[scale=0.60]{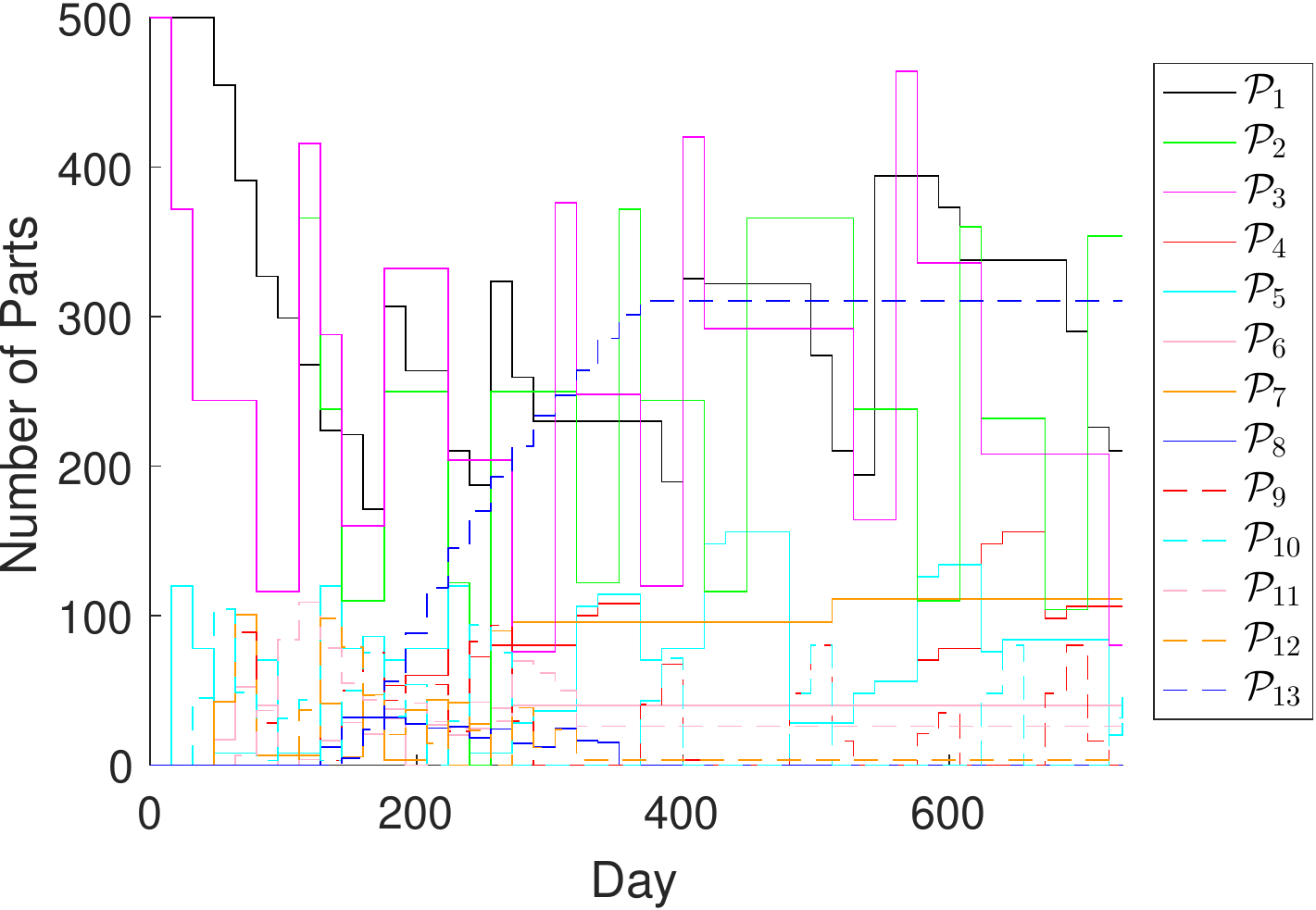}
\includegraphics[scale=0.60]{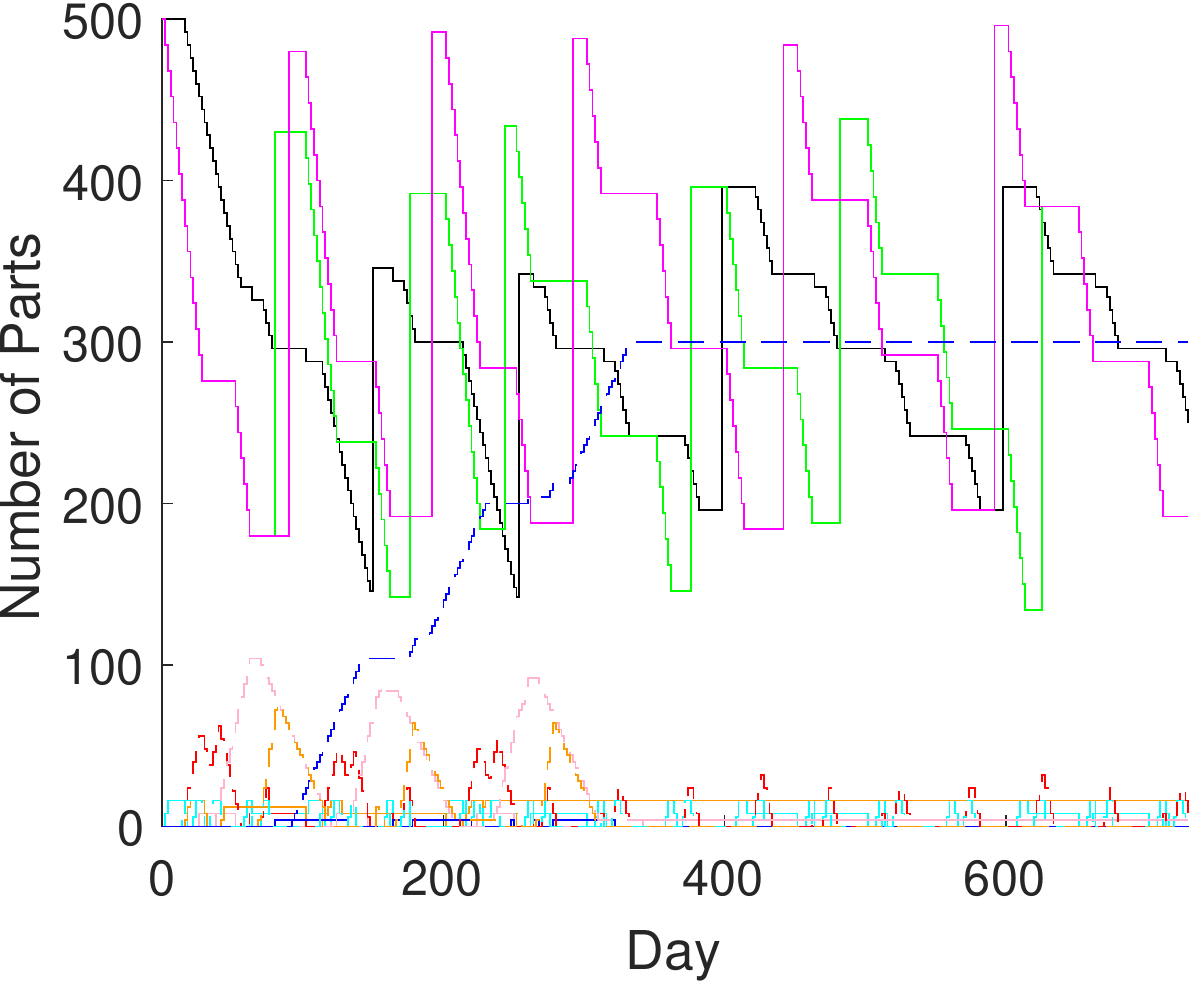}
\caption{The time history of the state vector in the complex pull system with transportation. From left to right, the trajectories are simulated using time buckets $\Delta t = 16$ days and $\Delta t = 2$ days, respectively.}\label{fig:timehistory_mix_transport}
\end{figure}

Figure \ref{fig:timehistory_mix_transport} shows the simulated numbers of parts in the system as they evolve in time using two different lengths of time buckets, i.e., $\Delta t = 16$ and $2$ days. The start of the delivery of the final product-${\cal{P}}_{13}$ has been shifted to a later date compared to that of the case without transportation. Thanks to the creation of the new processes, we are able to simulate the number of goods in the buffers right after their production, during the transportation and in the buffers before their instantaneous consumption in the following process. 

\begin{figure}[p]
\centering
\subcaptionbox{}{\includegraphics[scale=0.3]{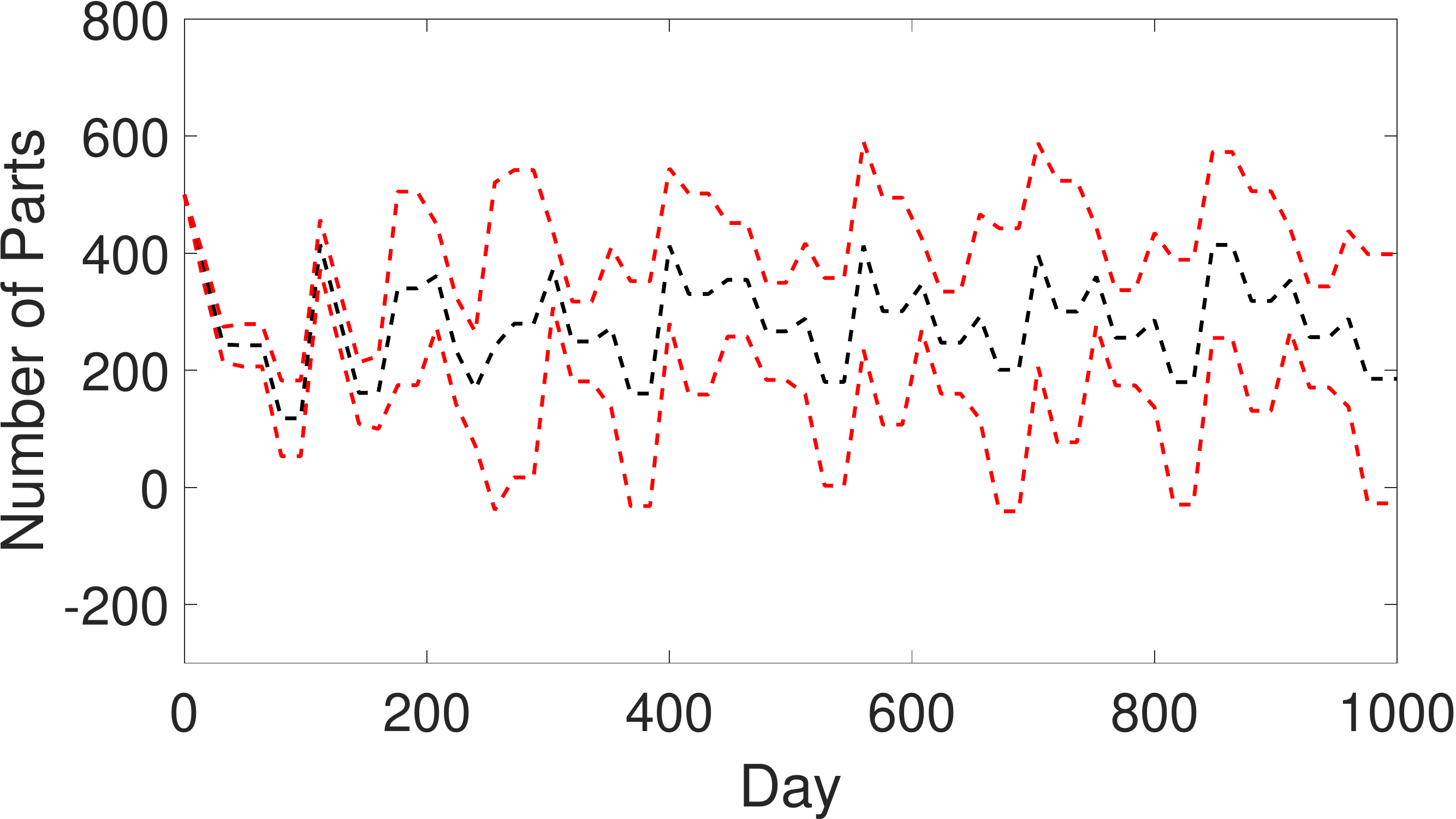}}
\subcaptionbox{}{\includegraphics[scale=0.3]{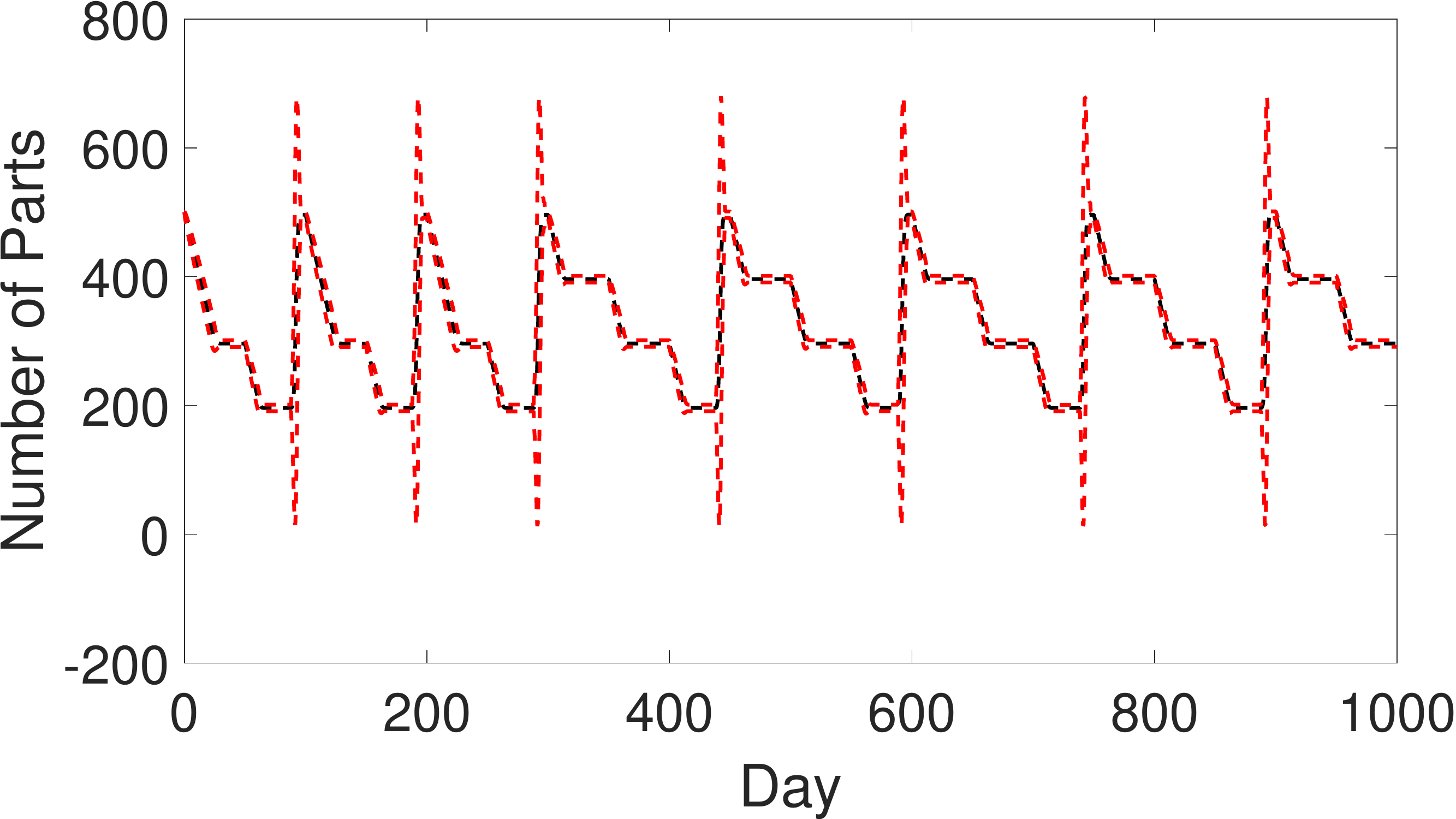}}
\vspace{-10pt}
\caption{ Stochastic pull system with both orders and transportation. (a) $\Delta t = 16$ days, (b) $\Delta t = 0.5$ days. } \label{fig:sto-dt16dt05-pull_BothOrd-pn1238}
\end{figure}

Additionally, we simulate a stochastic pull system with mixed orders and transportation using the L-leap method.
We present the average number and its 95\% confidence interval in $700$ days for part ${\cal{P}}_3$ in Figure \ref{fig:sto-dt16dt05-pull_BothOrd-pn1238}. It is noteworthy that very large uncertainties exist at the points where the inventory possibly gets refilled. 

%From this figure, it is clear that the system dynamic is complicated, but we still can observe that the number of peaks, which are given by the 95\%  confidence interval, does not change dramatically. Even though the shape of the curve for part  ${\cal{P}}_{13}$ looks different when we decrease $\Delta t$ from $16$ to $0.5$ days, the low resolution trend still can provide us an acceptable estimation of the final production time, i.e., around 400 days.

\subsection{Uncertainty propagation using MLMC - push system}\label{eg:uqpush}
We use MLMC to compute the expected number of ${\cal{P}}_8$ delivered in $300$ days in the previous push system. We consider $13$ random parameters, i.e., $\lambda_1 $-$\lambda_5 $ are the production rates of the processes $1$-$5$, $x_1(t=0)$, $x_2(t=0)$, $x_3(t=0)$ are the initial inventories of ${\cal{P}}_1$-${\cal{P}}_3$, $\hat{t}_i, i=1,\dots , 5$ are the processing time of processes $1$-$5$. The parameters are independently uniformly distributed as follows:

\begin{align}
&\lambda_1 \sim {\cal{U}}(8, 12)\,,\quad \lambda_2 \sim {\cal{U}}(8, 12)\,,\quad \lambda_3 \sim {\cal{U}}(4, 6)\,,\quad
\lambda_4 \sim {\cal{U}}(8, 12)\,,\quad \lambda_5 \sim {\cal{U}}(1, 3)\,, \nonumber\\
&x_1(t=0) \sim {\cal{U}}(800, 1200)\,,\quad x_2(t=0) \sim {\cal{U}}(300, 700)\,,\quad x_3(t=0) \sim {\cal{U}}(800, 1200)\,,\nonumber\\
&\hat{t}_1 \sim {\cal{U}}(1, 2)\,,\quad \hat{t}_2 \sim {\cal{U}}(1, 2)\,,\quad \hat{t}_3 \sim {\cal{U}}(10, 20)\,,\quad
\hat{t}_4 \sim {\cal{U}}(1, 2)\,,\quad \hat{t}_5 \sim {\cal{U}}(10, 50)\,. \nonumber
\end{align}

We evenly split the total tolerance between the bias and statistical error, i.e.,
$\epsilon^2_b = 0.5TOL^2 \quad \bar{\epsilon}_s^2 = 0.5TOL^2\,. $ The estimated values of $a$, $b$ and $g$ are $1.5$, $2$ and $1$, respectively. We show the numbers of samples in Table \ref{tb:nos}. The number of required levels commonly increases as we decrease the tolerance. Nevertheless, note that we have over-killed the bias when the tolerances are set to be $2.5\%$ and $0.5\%$, consequently, the number of levels does not change when the tolerance decreases to $1.25\%$ and $0.25\%$ respectively. 

\begin{table}[p]
\begin{center}
  \begin{tabular}{c | c | c | c | c | c | c | c | c | c }
    \hline
    Tol (percentage) & level 0 & level 1 & level 2 & level 3 & level 4 & level 5 & level 6 &level 7 &level 8 \\ \hline
1 (0.25\%) &    48610 & 13071 & 4248 & 1720 & 617 & 189 & 61 & 28 & 15 \\ \hline
2 (0.5\%) &    11872 & 3110 &  1085 & 441 & 164 & 53 & 16 & 6 & 2 \\ \hline
5 (1.25\%) &    1700 & 459 & 145 & 58 & 21 & 7 \\ \hline
10 (2.5\%) &    376 & 103 & 35 & 14 & 5 & 2 \\ \hline
30 (7.5\%) &    43 & 15 & 5 & 2 & 1 \\ \hline
  \end{tabular}
  \caption{Number of samples associated with different levels in MLMC given the tolerance.} \label{tb:nos}
\end{center}
\end{table}

The results and costs of the MLMC estimator are listed in Table \ref{tb:results}. It is shown that we achieved $3.27$ times acceleration when tolerance is 
$30 (7.5\%)$, $7$ times acceleration when the tolerances are $10 (2.5\%)$ and $5 (1.25\%)$, $70$ times acceleration when the tolerances are $2 (0.5\%)$ and $1 (0.25\%)$. 

\begin{table}[p]
\begin{center}
  \begin{tabular}{ c | c | c | c }
    \hline
    Tol (percentage) & result & MLMC cost (second) & MC cost (second)  \\ \hline
    1 (0.25\%) & 401 & 33.9 & 2470 \\ \hline
    2 (0.5\%) & 404 & 8.1 & 615 \\ \hline
    5 (1.25\%) & 402 & 1.1 & 7.5 \\ \hline
    10 (2.5\%) & 414 & 0.24 & 1.7 \\ \hline
    30 (7.5\%) & 396 & 0.034 & 0.11 \\ \hline
  \end{tabular}
  \caption{The results and cost of MLMC compared to standard MC. } \label{tb:results}
\end{center}
\end{table}

\subsection{Uncertainty propagation using MLMC - pull system}\label{eg:uqpull_t}
\begin{figure}[p]
\includegraphics[scale=0.6]{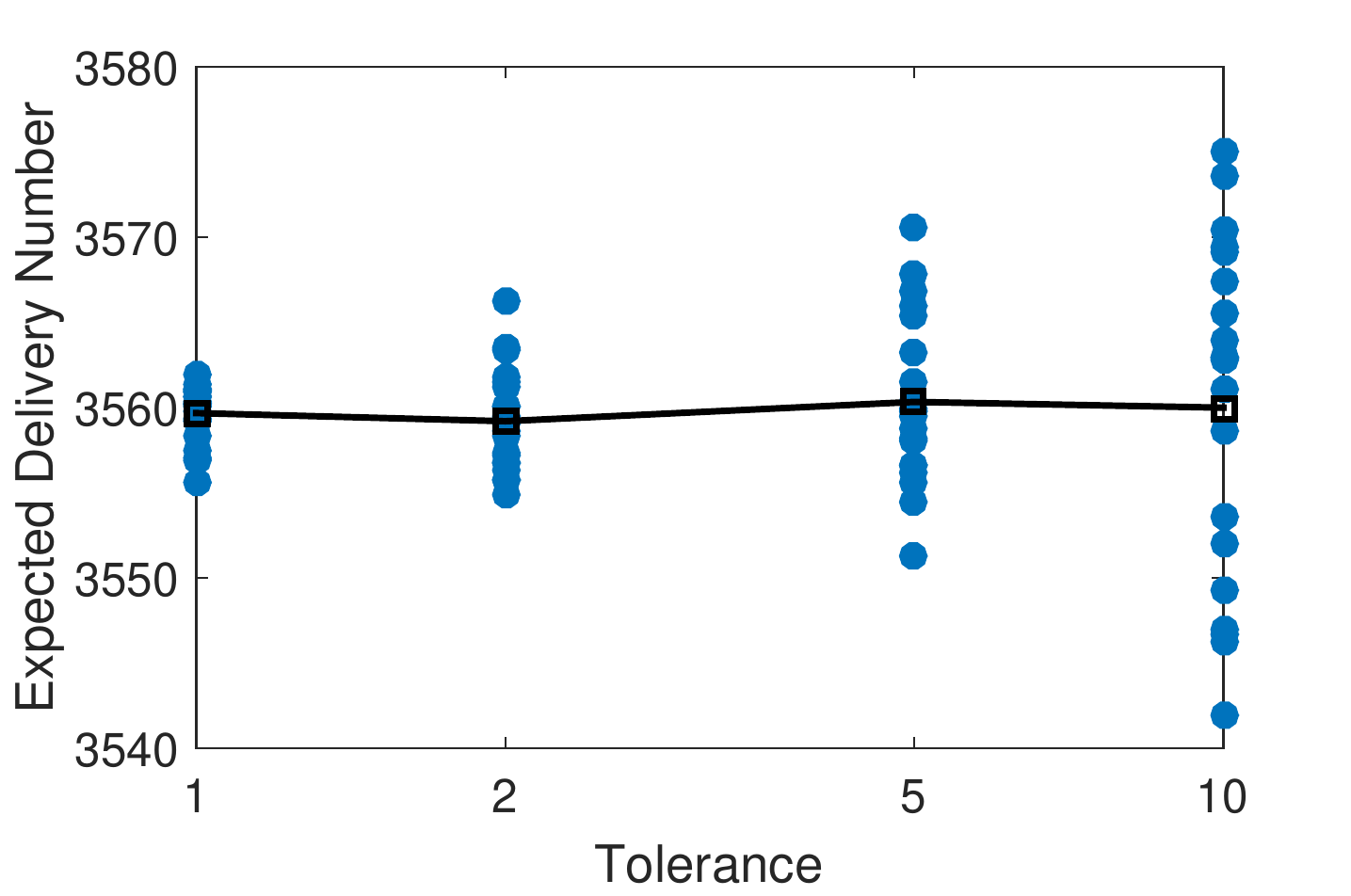}
\includegraphics[scale=0.6]{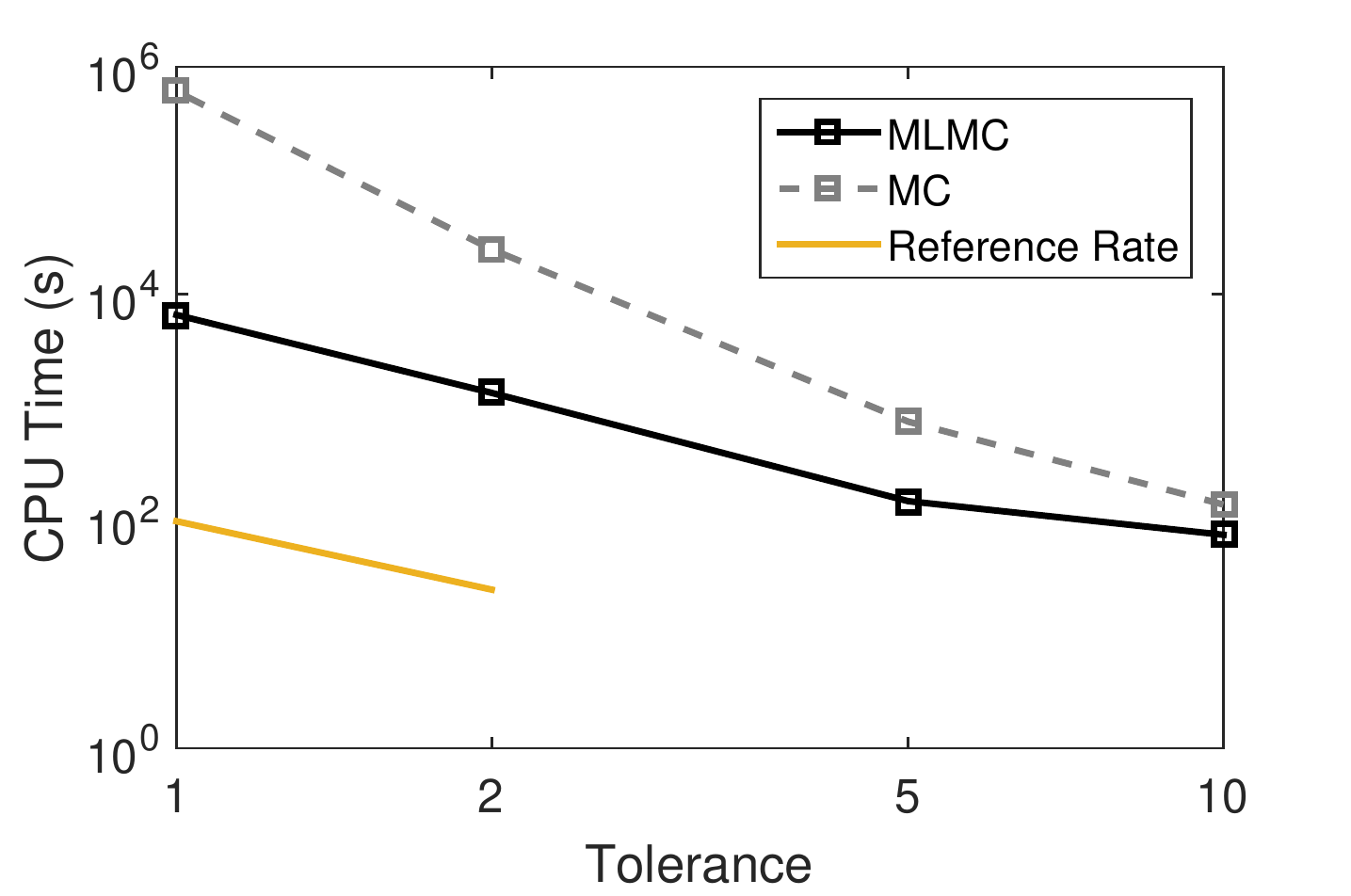}
\caption{The left figure shows the expected delivery of the final product in 3650 days in example \ref{eg:uqpull_t}; The right figure shows the average computational cost of MLMC w.r.t. the numerical tolerances in exmaple \ref{eg:uqpull_t}.The reference is computed using the MLMC theory in Section \ref{sec:UQ} ($a\approx 1.90$, $b\approx 0.78$, $g\approx 1.13$).}\label{fig:uqpull_quantity_p}
\end{figure}

In the last example, we consider both parametric and stochastic uncertainties for the pull system in \ref{eg:pull_t}. We vary in total $23$ parameters in the system. $\lambda_1$-$\lambda_5$ are the average production rates of the corresponding processes $1$-$5$, $\lambda_6$-$\lambda_{10}$ are the mean transportation rates associated with the processes $6$-$10$, $x_1(t=0)$, $x_2(t=0)$, $x_3(t=0)$ are the initial inventories of ${\cal{P}}_1$-${\cal{P}}_{3}$, $\hat{t}_i, i=1,\dots , 5$ are the processing time of processes $1$-$5$, $\hat{t}_i, i=6,\dots , 10$ are the transportation delays in the processes $6$-$10$. The parameters are independently uniformly distributed as follows:

\begin{align}
&\lambda_1 \sim {\cal{U}}(8, 12)\,,\quad \lambda_2 \sim {\cal{U}}(8, 12)\,,\quad \lambda_3 \sim {\cal{U}}(4, 6)\,,\quad
\lambda_4 \sim {\cal{U}}(8, 12)\,,\quad \lambda_5 \sim {\cal{U}}(1, 3)\,, \nonumber\\
&\lambda_{6} \sim {\cal{U}}(7, 9)\,,\quad \lambda_{7} \sim {\cal{U}}(7, 9)\,,\quad \lambda_{8} \sim {\cal{U}}(7, 9)\,,\quad
\lambda_{9} \sim {\cal{U}}(7, 9)\,,\quad \lambda_{10} \sim {\cal{U}}(1, 2)\,,\nonumber \\
&x_1(t=0) \sim {\cal{U}}(800, 1200)\,,\quad x_2(t=0) \sim {\cal{U}}(300, 700)\,,\quad x_3(t=0) \sim {\cal{U}}(800, 1200)\,,\nonumber\\
&\hat{t}_1 \sim {\cal{U}}(1, 2)\,,\quad \hat{t}_2 \sim {\cal{U}}(1, 2)\,,\quad \hat{t}_3 \sim {\cal{U}}(10, 20)\,,\quad
\hat{t}_4 \sim {\cal{U}}(1, 2)\,,\quad \hat{t}_5 \sim {\cal{U}}(10, 50)\,, \nonumber\\
&\hat{t}_6 \sim {\cal{U}}(8, 12)\,,\quad \hat{t}_7 \sim {\cal{U}}(8, 12)\,,\quad \hat{t}_8 \sim {\cal{U}}(8, 12)\,,\quad
\hat{t}_9 \sim {\cal{U}}(40, 60)\,,\quad \hat{t}_{10} \sim {\cal{U}}(8, 12)\,, \nonumber
\end{align}
We impose repetitive final product orders ($100$ quantities per order) with $100$ days' intervals. We also impose spare-part orders for parts $4$, $5$, $6$ and $7$: $30$ parts per order, every $30$ days.
We evenly split the total tolerance between the bias and statistical error, i.e.,
$\epsilon^2_b = 0.5TOL^2\,, \,\bar{\epsilon}_s^2 = 0.5TOL^2$. 
Under parametric uncertainties, the expected numbers of deliveries of the final products in $3650$ days are shown in the left picture of Figure \ref{fig:uqpull_quantity_p}, where the MLMC simulations are repeated $20$ times for four different values of the tolerances, i.e., $1$, $2$, $5$ and $10$. The mean value converges to $3560$ as we decrease the tolerance, while the variability of the estimator is tightly controlled by the prescribed tolerance.  The right picture of Figure \ref{fig:uqpull_quantity_p} compares the computational time of the MLMC with standard MC. It is note-worthy that the MLMC can be several magnitudes more efficient than standard MC as it has a much smaller rate of growth w.r.t. the tolerance than MC.

\begin{figure}[p]
\includegraphics[scale=0.6]{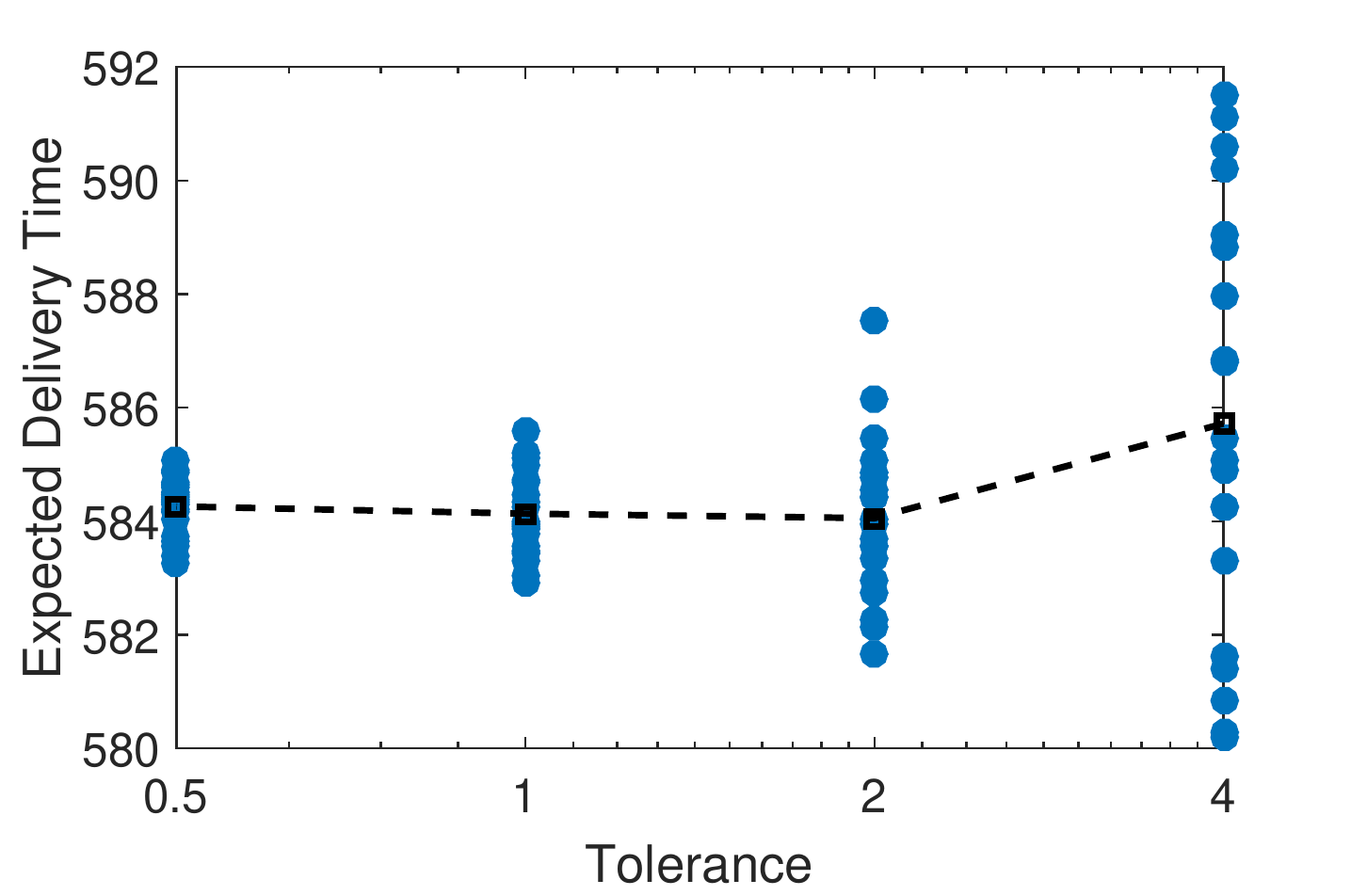}
\includegraphics[scale=0.6]{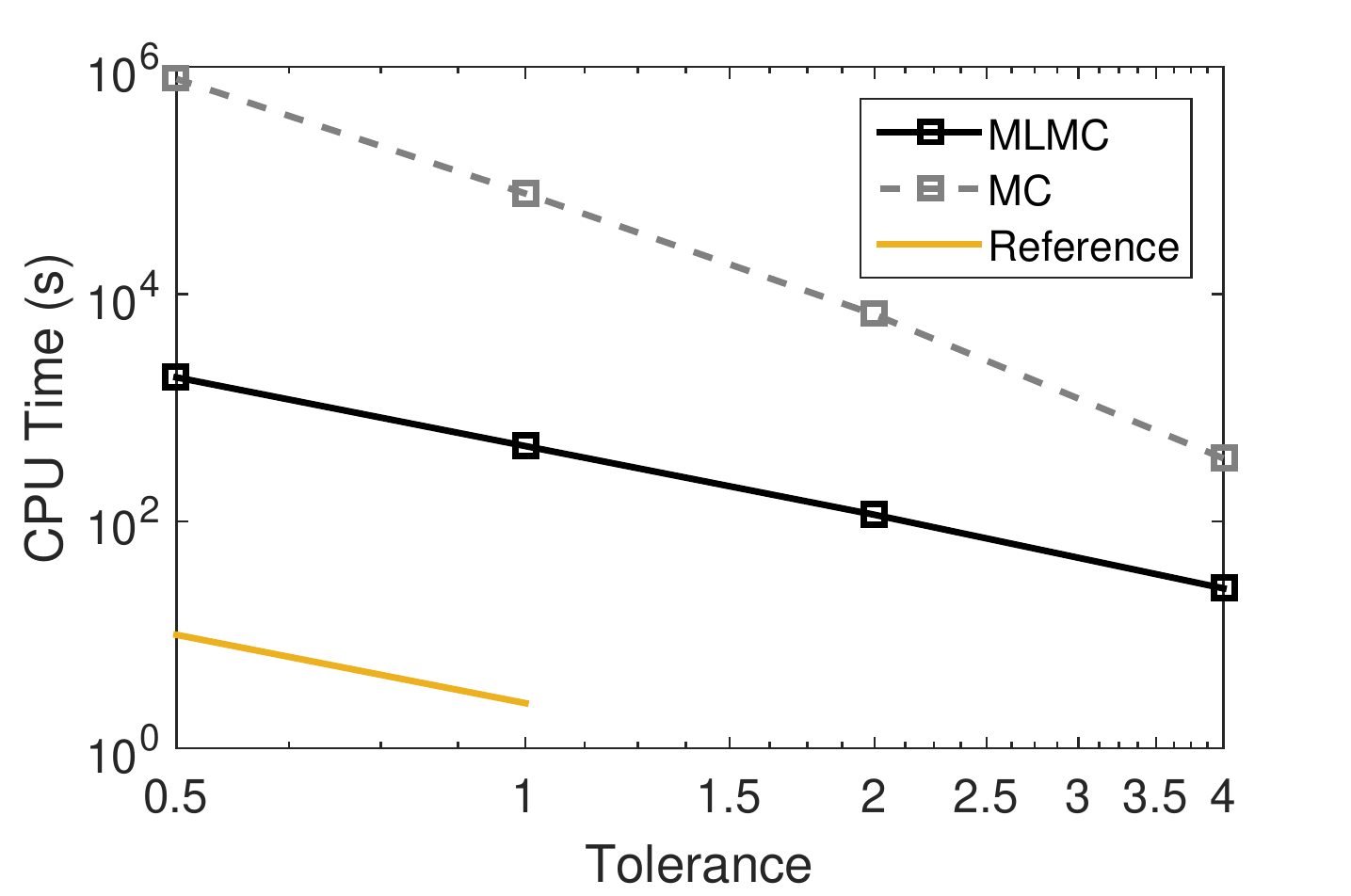}
\caption{The left figure shows the expected delivery time of $500$ products in example \ref{eg:uqpull_t}; The right figure shows the average computational cost of MLMC w.r.t. the numerical tolerances in example \ref{eg:uqpull_t}. The reference is computed using the MLMC theory in Section \ref{sec:UQ} ($a\approx 1.37$, $b\approx 1.44$, $g\approx 0.98$).}\label{fig:uqpush_time}
\end{figure}

We furthermore compute the expected delivery time of $500$ final products. The left picture in Figure \ref{fig:uqpush_time} shows $20$ batches of MLMC simulations of the delivery time for four different values of the tolerances, i.e., $0.5$, $1$, $2$ and $4$ days. The mean value converges to $584$ days, while we also observe that the variability of the MLMC results is controlled rigorously by the tolerance. The right picture of Figure \ref{fig:uqpush_time} compares the computational time of the MLMC estimator with the standard MC. Again, the MLMC is several magnitudes  faster than standard MC. More specifically, it is $10$ times faster than MC when the tolerance is $4$ days. This factor grows to $100$ as we reduce the tolerance to $0.5$ days. 

\begin{figure}[pt]
\includegraphics[scale=0.6]{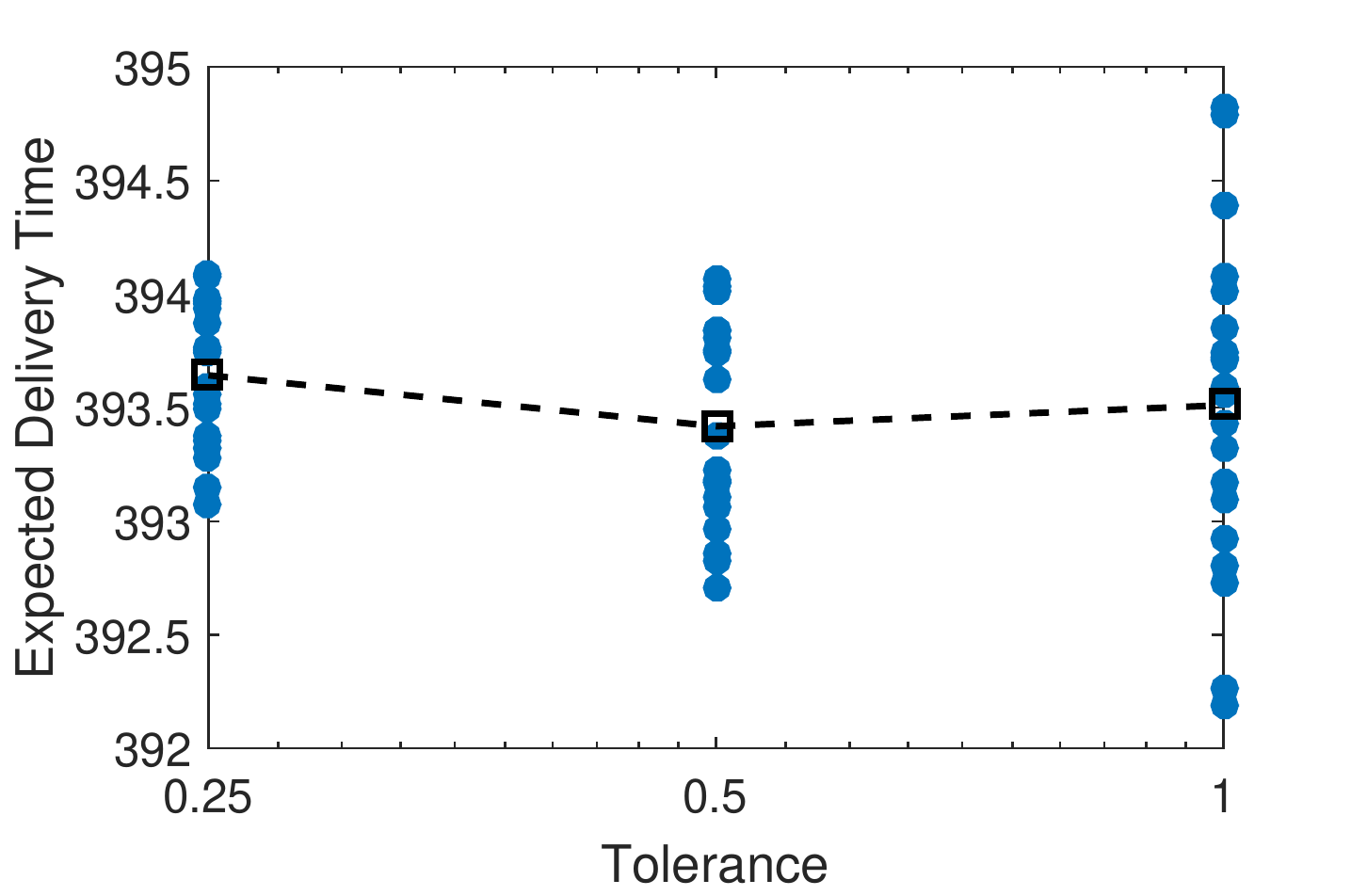}
\includegraphics[scale=0.6]{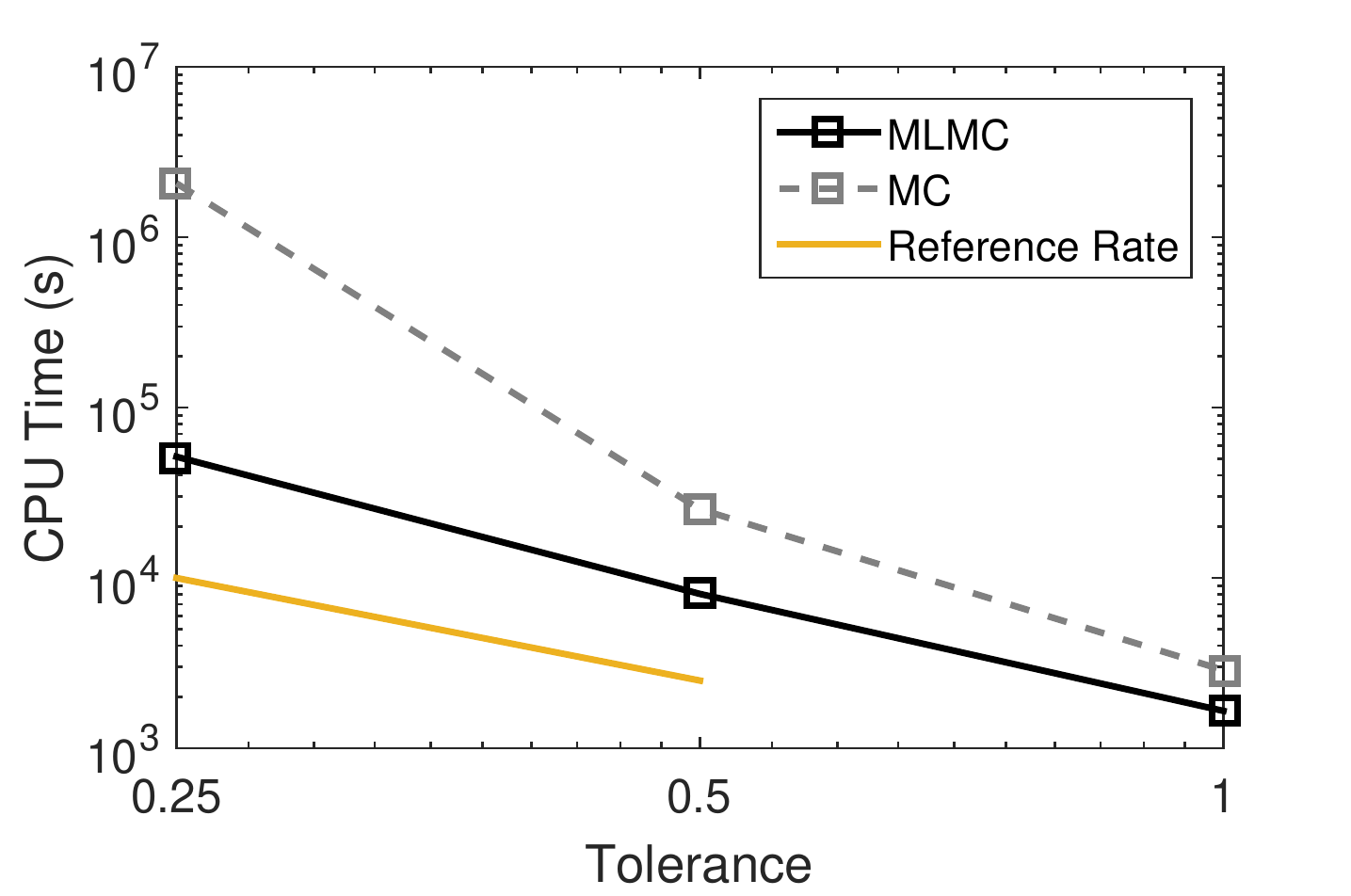}
\caption{The left figure shows the expected delivery time of $300$ final product in example \ref{eg:uqpull_t}; The right figure shows the average computational cost of MLMC w.r.t. the numerical tolerances in exmaple \ref{eg:uqpull_t}. The reference is computed using the MLMC theory in Section \ref{sec:UQ} ($a\approx 1.24$, $b\approx 1.07$, $g\approx 0.97$).}\label{fig:uqpull_quantity}
\end{figure}

Finally, we consider both parametric uncertainties and uncertainties driven by stochastic processes. Specifically, the numbers of processes happening in any time bucket is a Poisson random variable \eqref{eq:Delta_C_sto} and the number of production is an stochastic process related to a Binomial distribution \eqref{eq:Delta_P_sto}. We compute the expectation of the delivery time of $300$ final products using MLMC. We choose $\Delta t = 5$ days as the coarsest level. The left picture of Figure \ref{fig:uqpull_quantity} shows the results of $20$ runs of MLMC against the tolerances. The average delivery time converges to $393.6$ days. The right picture of Figure \ref{fig:uqpull_quantity} shows the average computational costs of the MLMC w.r.t. the tolerances. For tolerance smaller than $1$, the MLMC is significantly advantageous to the standard MC as the multilevel complexity grows much slower than the MC. The reference rate of MLMC's complexity, i.e., $\epsilon^{-2}$ ($a=1.24$, $b=1.07$, $g=0.97$), is similar to the growth of the measured CPU time. 

\section{Conclusion}\label{conclusion}
We had presented a multilevel uncertainty propagation framework utilizing time bucket method of simulating manufacturing supply chains. We incorporated several essential features for supply chain simulations, for example, limited capacities, push and pull productions, transportation, inventory refilling, and priority productions, into the leap methods which were previously used to approximate the DES of chemical and biochemical systems. The time buckets naturally offer a hierarchy of models which can be combined with MLMC to accelerate the propagation of uncertainties in a supply chain network. We demonstrated more than $10$ times speed up using our approach compared to standard MC using several manufacturing supply chain examples. Considering future work, we note that the framework of combining time buckets and MLMC can be applied to the agent-based \cite{swaminathan1998modeling} and continuous modeling \cite{d2010modeling} of supply chains to achieve efficient uncertainty propagation. 

\section{Acknowledgment}
The authors would like to acknowledge support from United Technologies Research Center through the innovation pipeline program and the capability program of the systems department. We thank Thomas Frewen and Bob Labarre for valuable proofreading. 

%\bibliographystyle{elsarticle-num}

%\begin{thebibliography}{00}
%%% Text of bibliographic item
%
%\bibitem{}
%
%\end{thebibliography}
\end{document}